\pdfoutput=1
\documentclass[a4paper,USenglish,cleveref,autoref,thm-restate,pdfa]{lipics-v2021}
\hideLIPIcs   %

\newif\ifanonymous
\newif\ifshort

\title{Proofs of Proof-of-Stake with Sublinear Complexity} %

\ifanonymous
\else%
\author{Shresth Agrawal}{Technische Universität München, Germany}{s.agrawal@tum.de}{https://orcid.org/0000-0002-7914-5979}{}%
\author{Joachim Neu}{Stanford University, USA}{jneu@stanford.edu}{https://orcid.org/0000-0002-9777-6168}{}%
\author{Ertem Nusret Tas}{Stanford University, USA}{nusret@stanford.edu}{https://orcid.org/0000-0001-6061-9700}{}%
\author{Dionysis Zindros}{Stanford University, USA}{dionyziz@stanford.edu}{https://orcid.org/0000-0002-1978-594X}{}%
\fi

\authorrunning{S. Agrawal, J. Neu, E. N. Tas, and D. Zindros}
\Copyright{Shresth Agrawal, Joachim Neu, Ertem Nusret Tas, and Dionysis Zindros}



\category{} %

\relatedversion{} %

\ifanonymous{}\else{%
\funding{JN is supported by the Protocol Labs PhD Fellowship.
ENT is supported by the Stanford Center for Blockchain Research.
The work of DZ was supported in part by funding from Harmony.}%
}\fi

\ifanonymous{}\else{%
\acknowledgements{The authors thank Kostis Karantias for the helpful discussions on bisection games,
and Daniel Marin for reading early versions of this paper and providing suggestions.
The work of JN
was conducted in part
while at Paradigm. 
The work of SA was conducted in part while at Common Prefix.}%
}\fi

\nolinenumbers %

\usepackage{preamble}

\allowdisplaybreaks
\EventEditors{Joseph Bonneau and S. Matthew Weinberg}
\EventNoEds{2}
\EventLongTitle{5th Conference on Advances in Financial Technologies (AFT 2023)}
\EventShortTitle{AFT 2023}
\EventAcronym{AFT}
\EventYear{2023}
\EventDate{October 23-25, 2023}
\EventLocation{Princeton, NJ, USA}
\EventLogo{}
\SeriesVolume{282}
\ArticleNo{14}

\begin{document}
\maketitle

\begin{abstract}
  Popular Ethereum wallets (like MetaMask)
  entrust centralized infrastructure providers (\eg, Infura)
  to run the consensus client logic on their behalf.
  As a result, these wallets are light-weight
  and high-performant, but come with security risks.
  A malicious provider can mislead the wallet by
  faking payments and balances, or censoring transactions.
  On the other hand, light clients, which
  are not in popular use today, allow decentralization, but are
  concretely
  inefficient, often with asymptotically \emph{linear} bootstrapping complexity.
  This poses a dilemma between decentralization and performance.

  We design, implement, and evaluate
  a new proof-of-stake (PoS) \emph{superlight} client with
  concretely efficient and asymptotically
  \emph{logarithmic} bootstrapping complexity.
  Our proofs of proof-of-stake (PoPoS)
  take the form of a Merkle tree of PoS epochs.
  The verifier enrolls the provers in a bisection game,
  in which honest provers are destined to win once an adversarial
  Merkle tree is challenged at sufficient depth.
  We provide an implementation for mainnet Ethereum:
  compared to the state-of-the-art light client construction
  of Ethereum, our client improves
  time-to-completion by $9\times$,
  communication by $180\times$,
  and energy usage by $30\times$
  (when bootstrapping after $10$ years of consensus execution).
  As an important additional application,
  our construction can be used to realize trustless cross-chain bridges,
  in which the superlight client runs within a smart contract and
  takes the role of an on-chain verifier.
  We prove our construction is secure
  and show how to employ it for
  other PoS systems such as
  Cardano (with fully adaptive adversary), Algorand, and Snow White.
\end{abstract}

\section{Introduction}

If I want to check how much money I have on Ethereum, the most secure
way is to run my own full node. Sadly, this requires downloading more
than 500 GB of data and can take up to 5 days to sync. Because of this,
most Ethereum users today outsource the task of
maintaining the latest state to a third-party provider such as Infura
or Alchemy. This allows the user to run a lightweight wallet, such as
MetaMask%
\footnote{MetaMask has $21{,}000{,}000$ monthly active users as of July 2022~\cite{metamask-mau-22}
and is the most popular non-custodial wallet~\cite{metamask-mau-21}.},
on their browser or smartphone.

What can go wrong if Infura is compromised or turns malicious?
As per the current wallet design, the wallet will blindly trust the provider and
therefore show incorrect data. This is enough to perform a double spend.
For example, imagine Bob wants to sell his car to Eve. Regrettably, Eve has compromised
Infura. Eve claims that she has paid Bob. Bob checks his MetaMask wallet and sees
an incoming and confirmed transaction from Eve.
Unfortunately, even though the wallet is non-custodial,
this transaction never happened, but is fraudulently reported by
Infura to Bob. Since Bob trusts his wallet, and his wallet trusts Infura,
he hands over the car keys to Eve. By the time Bob realizes he cannot use this
money, Eve has long disappeared with his car in Venice.
From the point of view of Infura, this is a huge liability. If Infura is compromised,
then all of its users are immediately compromised.

This creates a dilemma between good performance and security for the users.
This problem is not unique to Ethereum and appears in all proof-of-stake (PoS) systems.
In this paper, we resolve this dilemma by constructing a PoS blockchain
client which is both efficient and secure.
We solve this problem by constructing a protocol that allows efficiently verifying
the proof-of-stake blockchain without downloading the whole PoS. We call
such constructions succinct \emph{proofs of proof-of-stake}. These allow us to build
\emph{superlight clients}, clients whose communication complexity grows
sublinearly with the lifetime of the system.

\myparagraph{Contributions.}

\begin{enumerate}
      \item We give the first formal definition for succinct proof of proof-of-stake (PoPoS) protocols.
      \item We put forth a solution to the long-standing problem of efficient PoS bootstrapping.
            Our solution is exponentially better than previous work.
      \item We report on our implementation of a fully functional
            and highly performant superlight client
            for mainnet Ethereum. It
            is the first such construction
            for Ethereum. We measure and contrast the performance of our client against
            the currently proposed design for Ethereum.
      \item We theoretically show our construction is secure for Ethereum and
            other PoS blockchains.
\end{enumerate}

\subsection{Construction Overview}
Let us discuss the intuition about how our construction works.
We start with the existing full node design and iteratively make
it more lightweight.

\myparagraph{Full nodes.}
A \emph{full node} client boots holding only the genesis block and
connects to other full nodes (known as \emph{provers})
in order to synchronize to the latest tip.
The full node
downloads the entire chain block-by-block verifying each
transaction in the process.
This incurs high communication and computational complexity.
Once the client verifies the latest tip, it has calculated
the latest state and can answer user queries such as
\emph{``what is my balance?''}.
In order for the full node to
get to the correct tip, at least one connection to an honest
peer is required (this is known as the \emph{existential honesty}
assumption~\cite{backbone,backbone-new,varbackbone,pass-asynchronous}).

\myparagraph{Sync committees.}
Let's try to improve on the efficiency of the full node to make it a \emph{light client}.
PoS protocols typically proceed in \emph{epochs} during which the validator set is fixed.
In each epoch, a subset of validators is elected by the protocol as
the epoch \emph{committee}. This is a set of public keys.
The security of the protocol assumes that the majority~\cite{snowwhite,ouroboros,praos,genesis}
or super-majority~\cite{casper,pbft,tendermint-paper,hotstuff}
of the committee members are honest during the epoch.
The current committee signs the latest state.
Therefore, the client does not need to download all the blocks,
but instead only needs to download the latest state and verify the committee signatures
on it.
However, the stake changes hands in every epoch.
Hence, to perform the verification at some later epoch,
the client needs to keep track of the current committee.
To help the client in this endeavor,
the committee members of each epoch, while active, sign a \emph{handover} message inaugurating
the members of the new committee~\cite{pos-sidechains}.
This enables the light client to discover the latest committee by processing
a sequence of such handovers.

\myparagraph{\Optclient construction.}
Light clients like these already exist. Regrettably, they still need to download all
committees of the past to verify the current state.
In this paper we propose better solutions.
The first idea, which we call the \emph{\optclient construction},
is for the prover to work as follows:
For each committee, take its members, concatenate them all together,
and hash them into one hash.
The prover then sends this sequence of hashes (one for each committee)
to the client. Since the client is connected to at least one honest prover,
at least one of these provers will answer truthfully. If multiple provers
give conflicting claims to the client, all it needs to do is to find
which one is truthful. To do this, it compares the claims of all provers
pairwise. If two provers disagree, the client focuses on the first point
of disagreement in their hash sequences, and asks each prover to provide
the respective handover signatures to substantiate their claim.
Each prover reveals the committee attested by the hash at that point,
the previous committee and the associated handover messages.
Upon validating these messages, which can be done locally and efficiently,
the client identifies the truthful party, and accepts its state.
A lying
prover will not be able to provide such a handover for an invalid committee.
Once the client rejects the invalid committee claims, it will have
calculated the latest committee, and can proceed from there as usual.

\myparagraph{Superlight clients.}
Even though the complexity is concretely improved,
the sequence of committee hashes still grows linearly with the lifetime of the protocol.
To achieve sublinear complexity,
we improve the procedure to find
the first point of disagreement.
To this end,
our final PoPoS protocol requires
each prover to organize its claimed sequence of committees---one per epoch---into a Merkle tree~\cite{merkle}.
The roots of those trees are then sent over to the client, who compares them.
Upon detecting disagreement at the roots, the client asks the provers to reveal the
children of their respective roots.
By repeating this process recursively on the mismatching children,
it arrives at the \emph{first point of disagreement} between the claimed committee sequences,
in logarithmic number of steps.
After the first point of disagreement is found, the client
works similarly to the \optclient construction. This process achieves logarithmic communication.

\myparagraph{Bridges.}
Our PoPoS construction has two main applications:
\emph{Superlight} proof-of-stake clients that can bootstrap very efficiently,
and \emph{trustless bridges} that allow the passing of information
from one proof-of-stake chain to another. For bridging, we note that
a trustless bridge is nothing more than an on-chain superlight client.
It connects a source PoS blockchain
to any other destination blockchain. To do this, a smart contract on the destination chain,
which runs an implementation of our superlight client code (\eg, in Solidity),
is deployed. Whenever some information of
interest appears on the source chain, any helpful but untrusted relayer can
submit this information, together with the PoPoS proof to the smart contract. If the
information is inaccurate, another relayer challenges the first one by participating
in our interactive bisection game within a dispute period~\cite{pow-sidechains},
submitting one transaction for every round of
interaction of the PoPoS protocol. If both chains are PoS, the bridge can be made
bidirectional by running PoPoS superlight clients on both chains. To incentivize
on-chain participation, relayers who submit accurate information are rewarded, whereas
relayers who submit inaccurate information need to put up a collateral which is
slashed and is used to reward the challenger.

\subsection{Implementation Overview}
In addition to our theoretical contributions, we report on our open source implementation
(spanning about 8,200 lines of TypeScript code and 2,300 downloads from the community)
of our protocol for the Ethereum mainnet. Our
implementation is fully functional and supports all RPC queries used by typical wallets,
from simple payments to complex smart contract calls. It takes the form of a modular daemon that
augments any existing Ethereum wallet's functionality (such as MetaMask's) to be trustless,
without changing any user experience.
We perform measurements using the light client protocol currently proposed for Ethereum.
We find that this protocol,
while much more efficient than a full node, is likely insufficient to support communication-,
computation-, and battery-constrained devices such as browsers and mobile phones. Next,
we measure the performance of our implementation of an \emph{\optclient} for Ethereum
that achieves significant gains over the traditional light client.
We demonstrate this implementation is already feasible for resource-constrained devices.
Lastly, our implementation includes a series of experiments introducing a
\emph{superlight client} for Ethereum that achieves exponential asymptotic
gains over the optimistic light client. These gains constitute concrete
improvements over the optimistic light client when the
blockchain system is long-lived and has an execution history of a few years.
We compare all three clients in terms of communication (bandwidth and latency),
computation, and energy consumption.

\subsection{Related Work}

\begin{table*}[tb]
    \centering
    \caption{Comparison with previous works in terms of
        \emph{asymptotic} $\widetilde{\Theta}$ communication complexity,
        interactivity,
        and cryptographic model.
        Interactivity is the number of rounds, ignoring constants.
        Low communication and rounds of interactivity are preferable.
        $N$: number of epochs. $L$: number of blocks per epoch.
        Common prefix parameter $k$ is constant.}
    \setlength{\tabcolsep}{2pt}\footnotesize
    \begin{tabular}{lcccccccc}
        \toprule
         & \makecell[cc]{\textbf{SPV}\\\textbf{PoW}\\\cite{backbone}}
         & \makecell[cc]{\textbf{KLS}\\\cite{popow}}
         & \makecell[cc]{\textbf{FlyClient}\\\cite{flyclient}}
         & \makecell[cc]{\textbf{Superblocks}\\\cite{nipopows,logspace}}
         & \makecell[cc]{\textbf{Full}\\\textbf{PoS}\\\cite{ouroboros}}
         & \makecell[cc]{\textbf{Mithril}\\\cite{mithril}}
         & \makecell[cc]{\textbf{Coda}\\\cite{coda}}
         & \makecell[cc]{\textbf{PoPoS}\\(this work)}
        \\
        \midrule
        \textbf{Communication}
         & $NL$
         & $\log(NL)$
         & $\text{poly}\log(NL)$
         & $\log(NL)$
         & $NL$
         & $N + L$
         & $1$
         & $\log(N)$
        \\
        \textbf{Interactivity}
         & $1$
         & $\log(NL)$
         & $1$
         & $1$
         & $1$
         & $1$
         & $1$
         & $\log(N)$
        \\
        \textbf{Work/stake}
         & work
         & work
         & work
         & work
         & stake
         & stake
         & both
         & stake
        \\
        \textbf{Model}
         & RO
         & RO
         & RO
         & RO
         & standard
         & RO
         & CRS
         & standard
        \\
        \textbf{Primitives}
         & \makecell[tc]{hash}
         & \makecell[tc]{hash}
         & \makecell[tc]{hash}
         & \makecell[tc]{hash}
         & \makecell[tc]{hash, sig}
         & \makecell[tc]{hash,                \\sig, ZK}
         & \makecell[tc]{hash,                \\sig, ZK}
         & \makecell[tc]{hash, sig}
        \\
        \bottomrule
    \end{tabular}
    \label{tab.comparison}
\end{table*}

Table~\ref{tab.comparison} compares the current paper with related work.
Proof-of-work superlight clients have been explored in the interactive~\cite{popow}
and non-interactive~\cite{nipopows} setting using various constructions~\cite{logspace,compactsuperblocks,flyclient}.
Such constructions are backwards compatible~\cite{velvet,velvet-nipopows}
and have been deployed in practice~\cite{gas-efficient}. They have
also been used in the context of bridges~\cite{burn,sidechains,pow-sidechains,crosschain-sok}.
Several proof-of-stake-specific clients~\cite{pos-sidechains,mithril,eth-lightclient}
improve the efficiency of full nodes, but
require linear communication.
Chatzigiannis et al.~\cite{sok-light-clients} provide a survey of light clients.

Our construction is based on refereed proofs~\cite{refereed-computation,arbitrum,wallets,lazylight}.
PoPoS constructions can also be built via
(recursive) SNARKs/STARKs~\cite{coda,plumo}, some achieving constant communication.
However, these are clients for chains that were designed from the start
to be
proof-friendly.
Current attempts~\cite{zkbridge,succinct-labs} to retrofit popular PoS protocols
(\eg, Ethereum and Cosmos) with SNARK-based
proofs are expensive (annual prover operating-cost
of six to seven figures USD).\footnote{%
      For example,
      according to \cite[Section~6]{zkbridge}, proving consensus (of $128$ validators)
      on \emph{one} Cosmos block
      takes $18$ seconds on $32$ instances of Amazon AWS \texttt{c5.24xlarge}.
      (We are unable to independently reproduce this finding because the authors of
      zkBridge have not disclosed their code.)
      At Cosmos' block rate of $1$ block per second,
      that would require
      $18 \times 32$ continuously operating \texttt{c5.24xlarge} instances,
      costing annually
      \$12,967,488 on Amazon AWS (annual pricing, \texttt{us-east-1} region, June 2023),
      or \$1,749,600 on Hetzner's equivalents.
      Scaling this proportionally for Ethereum
      ($\times 4$ for sync committee size $512$, $/12$ for block rate $1$ block per $12$ seconds)
      yields an estimate of \$583,333 annually.%
}
Additionally, since \cite{zkbridge,succinct-labs}
do not include recursive proving/verification,
they
require the validator to remain online (to avoid linear overhead),
which may be a suitable assumption for bridges, but not for bootstrapping light clients
(\eg, intermittently online mobile wallets).
Unlike proof-based approaches, our protocol does not require changes in the PoS protocol,
and relies on simple primitives
such as hashes and signatures.
Our prover is stateless and adds only a few milliseconds of computation time
to an existing full node.
Our light client allows for bootstrapping from genesis with sublinear overhead.

Some clients obtain a \emph{checkpoint}~\cite{cosmos-whitepaper,weak-subjectivity}
on a recent block from a trusted source,
after which they download only a constant number of block headers
to identify the tip.
Our construction allows augmenting these clients so that they can
\emph{succinctly verify} the veracity of a checkpoint
without relying on a trusted third party.

\subsection{Outline}
We present our theoretical protocol in a \emph{generic} PoS framework, which typical proof-of-stake systems
fit into. We prove our protocol is secure if the underlying blockchain protocol satisfies certain simple and
straightforward \emph{axioms}. Many popular PoS blockchains can be made to fit within our axiomatic framework.
We define our desired primitive, the proof of proof-of-stake (PoPoS), together with the axioms required
from the underlying PoS protocol in Sec.~\ref{sec.primitive}. We iteratively build and present our construction in
Secs.~\ref{sec.sequences} and~\ref{sec.bisection}. We present the security claims in Sec.~\ref{sec.analysis}.
For concreteness, and because it is the most prominent PoS protocol, we give a concrete
construction of our protocol for Ethereum in Sec.~\ref{sec.pos-eth}. Ethereum is
the most widely adopted PoS protocol. Interestingly, Ethereum
directly satisfies our axiomatic framework and does not require any changes on the consensus layer at all.
The applicability of our framework to other PoS chains such as Ouroboros (Cardano), Algorand, and Snow White
are discussed in 
\ifshort
the full version of this paper~\cite{popos-eprint}.
\else
App.~\ref{sec.other}.
\fi

The description of our implementation and the relevant experimental measurements showcasing the
advantages of our implementation are presented in Sec.~\ref{sec.experiments}.

\section{Preliminaries}

\myparagraph{Proof-of-stake.}
Our protocols work in the proof-of-stake (PoS) setting. In a PoS protocol,
participants
transfer value and maintain a balance sheet of \emph{stake}, or
\emph{who owns what}, among each other. It is assumed that the \emph{majority of stake}
is honestly controlled at every point in time. The PoS protocol uses the current stake
\emph{distribution} to establish consensus. The exact mechanism by which consensus is
reached varies by PoS protocol. Our PoPoS protocol works for popular PoS flavours.

\myparagraph{Primitives.}
Participants in our PoS protocol transfer stake by \emph{signing}
transactions using a signature scheme~\cite{katz}. The public key associated
with each validator is known by everyone. The signatures are key-evolving,
and honest validators delete their
old keys after signing~\cite{key-evolving,praos}.\footnote{
    Instead of key-evolving signatures, Ethereum relies on a concept called
    \emph{weak subjectivity}~\cite{weak-subjectivity}. This alternative assumption can also
    be used in the place of key-evolving signatures to prevent posterior corruption attacks~\cite{long-range-survey}.}
We also use a collision resistant hash function. We highlight that it does not need to
be treated in the Random Oracle model, and no trusted
setup is required for our protocol (beyond what the underlying PoS protocol may need).

\myparagraph{Types of nodes.}
The stakeholders who participate in maintaining the system's consensus are known as
\emph{validators}. In addition to those, other parties, who do not participate in maintaining
consensus, can join the system, download its full history, and discover its current
state. These are known as \emph{full nodes}. Clients that are interested in joining the
system and learning a small part of the system state (such as their user's balance)
without downloading everything are
known as \emph{light clients}. Both full nodes and light clients can join the system
at a later time, after it has already been executing for some duration $|\chain|$.
A late-joining light
client or full node must \emph{bootstrap} by downloading some data from its peers. The
amount of data the light client downloads to complete the bootstrapping process is
known as its \emph{communication complexity}. A light client is \emph{succinct} if its
communication complexity is $\mathcal{O}(\text{poly}\log(|\chain|))$ in the lifetime
$|\chain|$ of the system. Succinct light clients are also called \emph{superlight clients}.
The goal of this paper is to develop a PoS superlight client.

\myparagraph{Time.}
The protocol execution proceeds in discrete \emph{epochs}, roughly corresponding to
moderate time intervals such as one day. Epochs are further subdivided into \emph{rounds},
which correspond to shorter time durations during which a message sent by one honest
party is received by all others. In our analysis, we assume synchronous communication.
The validator set stays fixed during an epoch, and it is known one epoch in advance.
The validator set of an epoch is determined by the snapshot of stake distribution
at the beginning of the previous epoch.
To guarantee an honest majority of validators at any epoch,
we assume a \emph{delayed honest majority} for a duration of
\emph{two epochs}:
Specifically, if a snapshot of the current stake distribution is taken at the beginning
of an epoch, this snapshot satisfies the honest majority assumption for a duration of
two full epochs.
Additionally, we assume that the adversary is \emph{slowly adaptive}:
She can corrupt any honest party, while respecting the honest majority assumption,
but that corruption only takes place two epochs later.
This assumption will be critical in our construction of \emph{handover}
messages that allow members of one epoch to inaugurate a committee representing
the next epoch (\cf Sec.~\ref{sec.sequences}).

\myparagraph{The prover/verifier model.}
The bootstrapping process begins with a light client connecting to its full node peers
to begin synchronizing. During the synchronization process, the full nodes are trying
to convince the light client of the system's state. In this context, the light client
is known as the \emph{verifier} and the full nodes are known as the \emph{provers}.
As usual, we assume the verifier is connected to at least one honest prover.
The verifier queries the provers about the state of the system,
and can exchange multiple messages to interrogate them about the truth of their claims
during an \emph{interactive protocol}.

\myparagraph{Assumptions.}
We make two central assumptions:
Firstly, that the light client can communicate \emph{interactively} with full nodes. This is
contrary to, for example, proof-based clients. Interactivity
incurs a penalty when our light client runs on-chain,
because it requires receiving data over the course of multiple transactions. This
is expensive in gas and time consuming in delays.
Secondly, that the light client has \emph{at least one honest} connection.
Many protocols assume this.
For example, a Bitcoin full node is not secure if all connections are dishonest.
In current systems, such as Ethereum, light clients typically connect
to RPC nodes instead of full nodes. It is better to
trust at least \emph{one among many} RPC connections is honest (as opposed to having
a single RPC connection), but one may still
become eclipsed. To resolve this concern, we advocate for light clients to
connect to full nodes directly,
which would make this assumption more
reasonable. Due to these two assumptions, we caution the reader to be
aware of the current limitations of our work, understanding it is not always
applicable.

\myparagraph{Notation.}
We use $\epsilon$ and $[\,]$ to mean the empty string and empty sequence.
By $x \concat y$, we mean the string concatenation
of $x$ and $y$ encoded in a way that $x$ and $y$ can be unambiguously retrieved.
We denote by $|\chain|$ the length of the sequence $\chain$; by $\chain[i]$ the $i^\text{th}$ (zero-based)
element of the sequence, and by $\chain[-i]$ the $i^\text{th}$ element from the end.
We use $\chain[i{:}j]$ to mean the subarray of $\chain$ from the $i^\text{th}$ element
(inclusive) to the $j^\text{th}$ element (exclusive). Omitting $i$ takes the sequence
to the beginning, and omitting $j$ takes the sequence to the end. We write 
$A \preccurlyeq B$ to mean that the sequence $A$ is a prefix of $B$. 
We use $\lambda$ to denote the security parameter.
Following Go notation, in our multi-party algorithms, we use $m \dashrightarrow A$ to
indicate that message $m$ is
sent to party $A$ and $m \dashleftarrow A$ to indicate that message $m$ is
received from party $A$.

\myparagraph{Ledgers.}
The consensus protocol attempts to maintain a unified view of a \emph{ledger} $\ledger$.
The ledger is a sequence of \emph{transactions} $\ledger = (\tx_1, \tx_2, \ldots)$.
Each validator and full node has a different view of the ledger. We denote the ledger
of party $P$ at round $r$ as $\ledger^P_r$.
Nodes joining the protocol, whether they are validators, full nodes, or (super)light clients,
can also \emph{write} to the ledger by asking for a transaction to be included.
In a secure consensus protocol, all honestly adopted ledgers
are prefixes of one another. We denote the longest among these ledgers as $\ledger^\cup_r$,
and the shortest among them as $\ledger^\cap_r$.
We will build our protocol on top of PoS protocols that are secure. A \emph{secure} consensus protocol
enjoys the following two virtues:

\begin{definition}[Consensus Security]
    A consensus protocol is \emph{secure} if it is:

    \begin{enumerate}
        \item \textbf{Safe:}
              For any honest parties $P_1, P_2$ and rounds $r_1 \leq r_2$: $\ledger^{P_1}_{r_1} \preccurlyeq \ledger^{P_2}_{r_2}$.

        \item \textbf{Live:}
              If all honest validators attempt to \emph{write}
              a transaction during $u$ consecutive rounds $r_1, \ldots, r_u$, it is included in $\ledger^P_{r_u}$
              of any honest party $P$.
    \end{enumerate}
\end{definition}

\myparagraph{Transactions.}
A transaction encodes an update to the system's state.
For example, a transaction could indicate a value transfer of 5 units
from Alice to Bob. Different systems use different transaction formats, but the particular
format is unimportant for our purposes. A transaction can be applied on the
current \emph{state} of the system to reach a new state. Given a state $\st$ and a transaction
$\tx$, the new state is computed by applying the state transition function $\transition$ to the
state and transaction. The new state is then $\st' = \transition(\st, \tx)$.
For example, in Ethereum, the state of the system encodes a list of balances of all
participants~\cite{buterin,wood}. The system begins its lifetime by starting at
a genesis state $\st_0$. A ledger also corresponds to a particular system state,
the state obtained by applying its transactions iteratively to the genesis state.
Consider a ledger $\ledger = (\tx_1 \cdots \tx_n)$. Then
the state of the system is
$\transition(\cdots \transition(\genesisstate, \tx_1), \cdots, \tx_n)$.
We use the shorthand notation $\transition^*$ to apply a sequence of transactions
$\overline{\tx} = \tx_1 \cdots \tx_n$
to a state. Namely,
$\transition^*(\genesisstate, \overline{\tx}) = \transition(\cdots \transition(\genesisstate, \tx_1), \cdots, \tx_n)$.

Because the state of the system is large, the state is compressed using
an authenticated data structure (\eg, Merkle Tree~\cite{merkle}). We denote by
$\stc$ the state \emph{commitment}, which is this short representation of
the state $\st$ (\eg, Merkle Tree root).
Given a state commitment $\stc$ and a transaction $\tx$, it is
possible to calculate the state commitment $\left<\st'\right>$
to the new state $\st' = \transition(\st, \tx)$. However, this calculation may require
a small amount of auxiliary data $\pi$ such as a Merkle tree proof of inclusion
of certain elements in the state commitment $\stc$.
We denote the transition that is performed
at the state commitment level by the \emph{succinct transition function} $\left<\transition\right>$.
Concretely, we will write that
$\left<\transition(\st, \tx)\right> = \left<\transition\right>(\stc, \tx, \pi)$.
This means that, if we take state $\st$ and apply transaction $\tx$ to it using the
transition function $\transition$, and subsequently calculate its commitment
using the $\left<\cdot\right>$ operator, the resulting state commitment is the same
as the one obtained by applying the succinct transition function $\left<\transition\right>$
to the state commitment $\stc$ and transaction $\tx$
using the auxiliary data $\pi$. If the auxiliary data is incorrect, the function $\left<\transition\right>$
returns $\bot$ to indicate failure.
If the state commitment uses a secure authenticated
data structure such as a Merkle tree,
we can only find a unique $\pi$ that makes the $\left<\transition\right>$
function run successfully.

\section{The PoPoS Primitive}\label{sec.primitive}

\myparagraph{The PoPoS abstraction.}
Every verifier $\client$ online at some round $r$ holds a state commitment $\stc^\client_r$.
To learn about this recent state, the verifier connects to
provers $\mathcal{P} = \{P_1, P_2, \cdots, P_q\}$.
All provers except one honest party can be controlled by the adversary,
and the verifier does not know which party among the provers is honest
(the verifier is assumed to be honest).
The honest provers are always online. Each of them maintains a ledger $\ledger_i$. They are consistent
by the safety of the underlying PoS protocol.
Upon receiving a query from the verifier, each honest prover sends back a state commitment corresponding
to its current ledger.
However, the adversarial provers might provide incorrect or outdated commitments that are different
from those served by their honest peers.
To identify the correct commitment, the light client mediates an \emph{interactive} protocol
among the provers:
\begin{definition}[Proof of Proof-of-Stake]
  A \emph{Proof of Proof-of-Stake protocol} (PoPoS) for a PoS consensus protocol
  is a pair of interactive probabilistic polynomial-time
  algorithms $(P, V)$. The algorithm $P$ is the \emph{honest prover} and the algorithm $V$
  is the \emph{honest verifier}. The algorithm $P$ is ran by an online full node,
  while $V$ is a light client booting up for the first time
  holding only the genesis state commitment $\left<\genesisstate\right>$.
  The protocol is executed between $V$ and a
  set of provers $\mathcal{P}$.
  After completing the interaction, $V$ returns a state commitment $\stc$.
\end{definition}

\myparagraph{Security of the PoPoS protocol.}
The goal of the verifier is to output a state commitment consistent with the view of the honest provers.
This is reflected by the following security definition of the PoPoS protocol.

\begin{definition}[State Security]
  \label{def:state-security}
  Consider a PoPoS protocol $(P, V)$ executed at round $r$, where $V$ returns $\stc$.
  It is \emph{secure} with \emph{parameter} $\nu$ if
  there exists a ledger $\ledger$ such that $\stc = \transition^*(\genesisstate, \ledger)$,
  and $\ledger$ satisfies:

  \begin{itemize}
    \item \textbf{Safety:} For all rounds $r' \geq r + \nu$: $\ledger \preccurlyeq \ledger^{\cup}_{r'}$.
    \item \textbf{Liveness:} For all rounds $r' \leq r  - \nu$: $\ledger^{\cap}_{r'} \preccurlyeq \ledger$.
  \end{itemize}
\end{definition}

State security implies that the commitment returned by a verifier corresponds
to a state recently obtained by the honest provers.

\section{The \OptClient}
\label{sec.sequences}

Before we present our succinct PoPoS protocol, we introduce sync committees and
handover messages, two necessary components
we use
in our construction.
We also propose a highly performant \optclient
as a building block for the superlight clients.

\myparagraph{Sync committees.}
To allow the verifier to achieve state security, we introduce
a \emph{sync committee} (first proposed in the context of PoS
sidechains~\cite{pos-sidechains}).
Each committee is elected for the duration of an epoch,
and contains a subset, of fixed size $m$, of the public keys of the validators associated with that epoch.
The committee of the next epoch is determined in advance at the beginning
of the previous epoch.
All honest validators agree on this committee.
The validators in the sync committee are sampled
from the validator set of the corresponding epoch in such a manner
that the committee retains honest majority during the epoch.
The exact means of sampling
are dependent on the PoS implementation. One way to construct
the sync committee is to sample uniformly at random from the
underlying stake distribution using the epoch randomness of the
PoS protocol~\cite{ouroboros,minimal_light_client}.
The first committee $S^0$ is recorded by the genesis state $\genesisstate$.
We denote the set of public keys of the sync committee assigned to epoch
$j \in \mathbb{N}$ by $S^j$,
and each committee member public key
within $S^j$ by $S^j_i$, $i \in \mathbb{N}$.

\myparagraph{Handover signatures.}
During each epoch $j$,
each honest committee member $S^j_i$ of epoch $j$ signs the tuple
$(j + 1, S^{j+1})$, where $j+1$ is the next epoch index and $S^{j+1}$ is the set
of all committee member public keys of epoch $j+1$.
We let $\sigma^j_i$ denote the signature of $S^j_i$ on the tuple $(j+1,S^{j+1})$.
This signature means that member
$S^j_i$ approves the inauguration of the next epoch committee.
We call those \emph{handover signatures}\footnote{Handover signatures between PoS epochs were
    introduced in the context of PoS sidechains~\cite{pos-sidechains}. Some practical
    blockchain systems already implement similar handover signatures~\cite{nearbridge,horizon}.},
as they signify that the previous epoch committee \emph{hands over} control to the next committee.
When epoch $j+1$ starts, the members of the committee $S^j$ assigned to epoch $j$ can no longer use their keys
to create handover signatures.\footnote{This assumption can be satisfied using key-evolving
    signatures~\cite{key-evolving,praos}, social consensus~\cite{weak-subjectivity}, or a static honest majority assumption.}
As soon as more than
$\frac{m}{2}$ members of $S^j$ have approved the inauguration of the next epoch committee,
the inauguration is ratified. This collection of signatures for the handover between
epoch $j$ and $j+1$ is denoted by $\Sigma^{j+1}$, and is called the \emph{handover proof}.
A \emph{succession} $\mathbb{S} = (\Sigma^1, \Sigma^2, \ldots, \Sigma^j)$
at an epoch $j$ is the sequence of all
handover proofs across an execution until the beginning of the epoch.

In addition to the handover signature,
at the beginning of each epoch,
honest committee members sign the \emph{state commitment} corresponding to
their ledger.
When the verifier learns the latest committee, these signatures enable him to find the current state commitment.

\myparagraph{A naive linear client.}
Consider a PoPoS protocol, where each honest prover
gives the verifier a state commitment and signatures on the commitment
from the latest sync committee $S^{N-1}$, where $N$ is the number of epochs (and $N-1$ is the last epoch).
To convince the verifier that $S^{N-1}$ is the correct latest committee,
each prover also shares the sync committees $S^0 \ldots S^{N-2}$ and the associated handover proofs in its view.
The verifier knows $S_0$ from the genesis state $\genesisstate$, and can verify
the committee members of the future epochs iteratively through the handover proofs.
Namely, upon obtaining the sync committee $S^j$,
the verifier accepts a committee $S^{j+1}$ as the correct committee assigned to epoch $j+1$,
if there are signatures on the tuple $(j+1, S^{j+1})$ from over half of the committee members in $S^j$.
Repeating the process above, the verifier can identify the correct committee
for the last epoch.
After identifying the latest sync committee, the verifier checks
if the state commitment provided by a prover is signed by over half of the committee members.
If so, he accepts the commitment.

It is straightforward to show that this strawman PoPoS protocol (which we abbreviate as \expLClegend) is secure (Def.~\ref{def:state-security})
under the following assumptions:
\begin{enumerate}
    \item The underlying PoS protocol satisfies safety and liveness.
    \item The majority of the sync committee members are honest.
\end{enumerate}
When all provers are adversarial, the verifier might not receive any state commitment from them.
Even though generally at least one prover is assumed to be honest,
the strawman protocol does not need this for the \emph{correctness} of the commitment
accepted by the verifier,
since the verifier validates each sync committee assigned to consecutive epochs,
and does not accept commitments not signed by over $\frac{m}{2}$ members of the
latest committee.

Regrettably, the strawman protocol is $\mathcal{O}(|\chain|)$ and not succinct:
To identify the lastest
sync committee, the verifier has to download each sync committee since the genesis block.
In the rest of this paper, we will improve this protocol to make it succinct.

\myparagraph{The \optclient (\expOLClegend).}
We now reduce the communication complexity of the verifier.
In this version of the protocol,
instead of sharing the sync committees $S^0 \ldots S^{N-2}$ and the associated handover proofs,
each honest prover $P$ sends a sequence of \emph{hashes}
$h^1 \ldots h^{N-1}$ corresponding to the sync committees $S^0 \ldots S^{N-1}$.
Subsequently, to prove the correctness of the state commitment, the prover $P$
reveals the latest sync committee $S^{N-1}$ assigned to epoch $N-1$ and the signatures by its members on the commitment.
Upon receiving the committee $S^{N-1}$, the verifier checks
if the hash of $S^{N-1}$ matches $h^{N-1}$, and validates the signatures on the commitment.

Unfortunately, an adversarial prover $P^*$ can claim an incorrect committee $S^{*,N-1}$,
whose hash $h^{*,N-1}$ disagrees with $h^{N-1}$ returned by $P$.
This implies a disagreement between the two hash sequences received from $P$ and $P^*$.
The verifier can exploit this discrepancy
to identify the truthful party that returned the correct committee.
Towards this goal, the verifier iterates over the two hash sequences, and finds the \emph{first point of disagreement}.
Let $j$ be the index of this point such that $h^j \neq h^{*,j}$ and
$h^{i} = h^{*,i}$ for all $i < j$.
The verifier then requests $P$ to reveal the committees $S^j$ and $S^{j-1}$
at the preimage of $h^j$ and $h^{j-1}$, and to supply
a handover proof $\Sigma^{j}$ for $S^{j-1}$ and $S^{j}$.
He also requests $P^*$ to reveal the committees $S^{*,j}$ and $S^{*,j-1}$ at the preimage of $h^{*,j}$ and $h^{*,j-1}$,
and to supply a handover proof $\Sigma^{*,j}$ for $S^{*,j-1}$ to $S^{*,j}$.
As $h^{j-1} = h^{*,j-1}$ by definition, the verifier is convinced that
the committees $S^{j-1}$ and $S^{*,j-1}$ revealed by $P$ and $P^*$
are the same.

Finally, the verifier checks whether the committees $S^{*,j}$ and $S^{j}$ were inaugurated by the
previous committee $S^{j-1}$ using the respective handover proofs $\Sigma^{j}$ and $\Sigma^{*,j}$.
Since $S^{j-1}$ contains over $\frac{m}{2}$ honest members that signed only the correct
committee $S^j$ assigned to epoch $j$, adversarial prover $P^*$ cannot create a handover proof
with sufficiently many signatures inaugurating $S^{*,j}$.
Hence, the handover from $S^{j-1}$ to $S^{*,j}$ will not
be ratified $\Sigma^{*,j}$, whereas the handover from $S^{j-1}$ to $S^j$ will
be ratified by $\Sigma^{j}$.
Consequently, the verifier will identify $P$ as the truthful party and accept its commitment.

In the protocol above,
security of the commitment obtained by the prover relies crucially on the existance of an honest prover.
Indeed, when all provers are adversarial, they can collectively return the \emph{same}
incorrect state commitment and the \emph{same} incorrect sync committee for the latest epoch.
They can then provide over $\frac{m}{2}$ signatures by this committee on the incorrect commitment.
In the absence of an honest prover to challenge the adversarial ones,
the verifier would believe in the validity of an incorrect commitment.

The \optclient reduces the communication load of sending over the whole sync committee sequence
by representing each committee with a constant size hash.
However, it is still $\mathcal{O}(|\chain|)$ as the verifier has to do a linear search
on the hashes returned by the two provers to identify the first point of disagreement.
To support a truly succinct verifier, we will next work towards an interactive PoPoS protocol based on bisection games.

\begin{figure*}
    \centering
    \includegraphics[width=\textwidth,keepaspectratio]{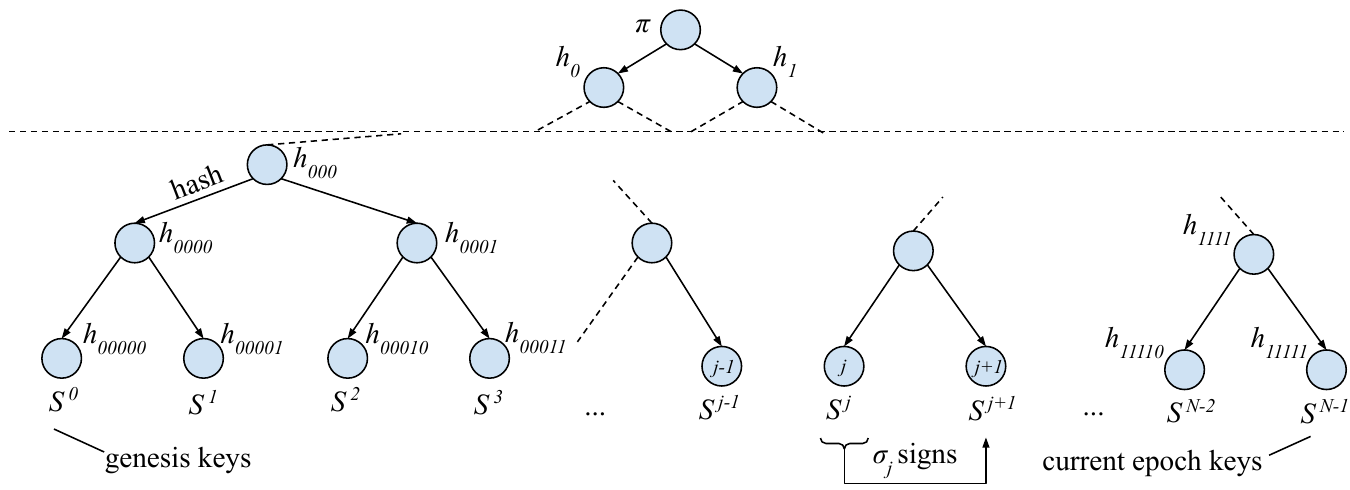}
    \caption{The \emph{handover tree}, the central construction of our protocol.
        The root of the Merkle tree is the initial proof $\pi$. During the bisection
        game, the signatures between the challenge node $j$ and its neighbours
        $j-1$ and $j+1$ are validated.}
    \label{fig.popos-tree}
\end{figure*}

\section{The Superlight Client}
\label{sec.bisection}

\myparagraph{Trees and mountain ranges.}
Before describing the succinct PoPoS protocol and the superlight client,
we introduce the data structures used by the bisection games.

Suppose the number of epochs $N$ is a power of two.
The honest provers organize the
committee sequences for the past epochs into a Merkle tree
called the \emph{handover tree} (Fig.~\ref{fig.popos-tree}).
The $j^\text{th}$ leaf of the handover tree contains the committee $S^j$ of the $j^\text{th}$ epoch.
A handover tree consisting of leaves $S^0, \ldots, S^{N-1}$
is said to be \emph{well-formed} with respect to a succession $\mathbb{S}$ if it satisfies the following properties:
\begin{enumerate}
    \item The leaves are syntactically valid. Every $j^\text{th}$ leaf
          contains a sync committee $S^j$ that consists of $m$ public keys.
    \item The first leaf corresponds to the known \emph{genesis} sync committee $S^0$.
    \item For each $j=1 \ldots N-1$, $\Sigma^j$ consists of over $\frac{m}{2}$ signatures
          by members of $S^{j-1}$ on $(j, S^j)$.
\end{enumerate}

Every honest prover holds a succession of handover signatures
attesting to the inauguration of each sync committee in its
handover tree after $S^0$. These successions might be
different for every honest prover as any set of signatures
larger than $\frac{m}{2}$ by $S^j$ can inaugurate $S^{j+1}$.
However, the trees are the same for all honest parties,
and they are well-formed with respect to the succession
held by each honest prover.

When the number $N$ of epochs is not a power of two,
provers arrange the past sync committees into Merkle mountain ranges (MMRs)~\cite{mmr,mmr-grin}.
An MMR is a list of Merkle trees, whose sizes are decreasing powers of two.
To build an MMR, a prover first obtains a binary representation $2^{q_1}+\ldots+2^{q_n}$ of $N$, where $q_1>\ldots>q_n$.
It then divides the sequence of sync committees into $n$ subsequences, one for each $q_i$.
For $i \geq 1$, the $i^\text{th}$ subsequence contains
the committees $S^{\sum_{n=1}^{i-1} 2^{q_i}}, \ldots, S^{(\sum_{n=1}^{i} 2^{q_i})-1}$.
Each $i^\text{th}$ subsequence is organized into a distinct Merkle tree $\mathcal{T}_i$, whose root,
denoted by $\left<\mathcal{T}_i\right>$, is called a \emph{peak}.
These peaks are all hashed together to obtain the root of the MMR.
We hereafter refer to the index of each leaf in these Merkle trees with the
epoch of the sync committee contained at the leaf.
(For instance, if there are two trees with sizes $4$ and $2$, the leaf indices in the first tree
are $0,1,2,3$ and the leaf indices in the second tree are $4$ and $5$.)
The MMR is said to be \emph{well formed} if each constituent tree is well-formed
(but, of course, only the first leaf of the first tree needs to contain the genesis committee).
To ensure succinctness, only the peaks and a small number of leaves,
with their respective inclusion proofs, will be presented to the verifier during the following bisection game.

\myparagraph{Different state commitments.}
We begin our construction of the full PoPoS protocol (abbreviated \expSLClegend for
\emph{Super Light Client})
by describing the first messages exchanged between
the provers $\mathcal{P}$ and the verifier.
Each honest prover first shares the state commitment signed by the latest sync committee
at the beginning of the last epoch $N-1$.
If all commitments received by the verifier are the same,
by existential honesty, the verifier rests assured this commitment is correct, \ie,
it corresponds to the ledger of the honest provers at the beginning of the epoch.
Otherwise, the verifier requests from each prover in $\mathcal{P}$:
(i) the MMR peaks $\mroot_i$, $i \in [n]$ held by the prover, where $n$ is the number of peaks,
(ii) the latest sync committee $S^{N-1}$,
(iii) a Merkle inclusion proof for $S^{N-1}$ with respect to the last peak $\mroot_{n}$, and
(iv) signatures by the committee members in $S^{N-1}$ on the state commitment given by the prover.

Upon receiving these messages,
the verifier first checks if there are more than $\frac{m}{2}$ valid signatures by the
committee members in $S^{N-1}$ on the state commitment.
It then verifies the Merkle proof for $S^{N-1}$ with respect to $\mroot_n$.
As the majority of the committee members in $S^{N-1}$ are honest, it is not possible
for different state commitments to be signed by over half of $S^{N-1}$.
Hence, if the checks above succeed for two provers $P$ and $P^*$ that returned different commitments,
one of them ($P^*$) must be an adversarial prover, and must have claimed
an incorrect sync committee $S^{*,N-1}$ for the last epoch.
Moreover, as the Merkle proofs for both $S^{*,N-1}$ and $S^{N-1}$
verify against the respective peaks $\mroot_n$ and $\mroot^*_n$,
these peaks must be different.
Since the two provers disagree on the roots and there is only one
well-formed MMR at any given epoch, therefore one of the provers does not hold a well-formed
MMR.
This reduces the problem of identifying the correct state commitment
to detecting the prover that has a well-formed MMR behind its peaks.

\begin{algorithm}[tb]
    \caption{Run by the verifier during the bisection game
        to identify the first point of disagreement between the provers' leaves
        (\cf Fig.~\ref{fig.bisection-game}).
        Here, $\hon{P}$ and $\adv{P^*}$ denote the honest and adversarial provers,
        whereas $\hon{\mroot}$ and $\adv{\mroot^*}$ denote the roots of their respective
        Merkle trees with size $\ell$.}
    \label{alg.bisection.verifier}
    \begin{algorithmic}[1]\small
        \Function{\sc FindDisagreement}{$\hon{P, \mroot}, \adv{P^*, \mroot^*}, \ell$}
            \State\Let{\hon{h_c}, \adv{h^*_c}}{\hon{\mroot}, \adv{\mroot^*}}\label{line:initial}
            \While{$\ell > 1$}
                \State\Receive{(\hon{h_0},\hon{h_1})}{\hon{P}}; \Receive{(\adv{h^*_0},\adv{h^*_1})}{\adv{P^*}}
                \If{$h_c \neq H(h_0 \concat h_1)$}\label{line:check1}
                    \Return\Comment{$P$ loses}
                \EndIf
                \If{$h^*_c \neq H(h^*_0 \concat h^*_1)$}\label{line:check2}
                    \Return\Comment{$P^*$ loses}
                \EndIf

                \If{$\hon{h_0} \neq \adv{h^*_0}$}\label{line:left}
                    \State\Let{\adv{h^*_c}, \hon{h_c}}{\adv{h^*_0}, \hon{h_0}};
                    \Send{(\textrm{open},0)}{\hon{P}}; \Send{(\textrm{open},0)}{\adv{P^*}}
                \Else\label{line:right}
                    \State\Let{\adv{h^*_c}, \hon{h_c}}{\adv{h^*_1}, \hon{h_1}};
                    \Send{(\textrm{open},1)}{\hon{P}}; \Send{(\textrm{open},1)}{\adv{P^*}}
                \EndIf
                \State\Let{\ell}{\ell//2}
            \EndWhile
            \State\Receive{\hon{S}}{\hon{P}}; \Receive{\adv{S^*}}{\adv{P^*}}
            \State\Return{$\hon{S},\adv{S^*}$}
        \EndFunction
    \end{algorithmic}
\end{algorithm}

\begin{algorithm}[tb]
    \caption{Run by an honest prover during the bisection game to reply to the verifier $V$'s queries. The sequence $\prover{S^0},\ldots,\prover{S^{N-1}}$ denotes the sync committees in the prover's view.}
    \label{alg.bisection.prover}
    \begin{algorithmic}[1]\small
        \Function{\sc ReplyToVerifier}{$\prover{S^0},\ldots,\prover{S^{N-1}}$}
            \State\Let{\prover{\mathcal{T}}}{\textsc{MakeMT}(\prover{S^0,\ldots,S^{N-1}})};
            \Send{\prover{\mathcal{T}}\textrm{.root}}{\verifier{V}};
            \Let{j}{0}
            \While{$\prover{\mathcal{T}}\textrm{.size} > 1$}
                \State\Send{(\prover{\mathcal{T}}\textrm{.left.root}, \prover{\mathcal{T}}\textrm{.right.root})}{\verifier{V}};\label{line:reveal}
                \Receive{(\text{open},\verifier{i})}{\verifier{V}}\label{line:comm}
                \If{$\verifier{i}=0$}
                    \State\Let{\prover{\mathcal{T}}}{\prover{\mathcal{T}}\textrm{.left}}\label{line:descend1}
                \Else
                    \State\Let{\prover{\mathcal{T}}}{\prover{\mathcal{T}}\textrm{.right}}\label{line:descend2}
                \EndIf
                \State\Let{j}{2j + \verifier{i}}
            \EndWhile
            \State\Send{\prover{S^j}}{\verifier{V}}
        \EndFunction
    \end{algorithmic}
\end{algorithm}

\myparagraph{Bisection game.}
To identify the honest prover with the well-formed MMR,
the verifier (Alg.~\ref{alg.bisection.verifier}) initiates a bisection game between $P$ and $P^*$ (Alg.~\ref{alg.bisection.prover}).
Suppose the number of epochs $N$ is a power of two.
Each of the two provers claims to hold a tree with size $N$
(otherwise, since the verifier knows $N$ by his local clock,
the prover with a different size Merkle tree loses the game.)
During the game, the verifier aims to locate the first point of disagreement
between the alleged sync committee sequences
at the leaves of the provers' Merkle trees, akin to the improved \optclient (Sec.~\ref{sec.sequences}).

The game proceeds in a binary search fashion similar
to refereed delegation of computation~\cite{refereed-computation,practical-delegation,arbitrum}.
Starting at the Merkle roots $\mroot$ and $\mroot^*$ of the two trees,
the verifier traverses an identical path on both trees
until reaching a leaf with the same index.
This leaf corresponds to the first point of disagreement.
At each step of the game, the verifier asks the provers to reveal the children of the
\emph{current} node,
denoted by $h_c$ and $h^*_c$ on the respective trees (Alg.~\ref{alg.bisection.prover}, l.~\ref{line:reveal}).
Initially, $h_c=\mroot$ and $h^*_c=\mroot^*$ (Alg.~\ref{alg.bisection.verifier}, l.~\ref{line:initial}).
Upon receiving the alleged left and right child nodes $h^*_0$ and $h^*_1$ from $P^*$,
and $h_0$, $h_1$ from $P$, he checks if $h_c=H(h_0 \concat h_1)$ and $h^*_c=H(h^*_0 \concat h^*_1)$,
where $H$ is the collision-resistant hash function used to construct the Merkle trees (Alg.~\ref{alg.bisection.verifier}, ll.~\ref{line:check1} and~\ref{line:check2}).
The verifier then compares $h_0$ with $h^*_0$, and $h_1$ with $h^*_1$
to determine if the disagreement is on the left or the right child (Alg.~\ref{alg.bisection.verifier}, ll.~\ref{line:left} and~\ref{line:right}).
Finally, he descends into the first disagreeing child, and communicates this decision to the provers (Alg.~\ref{alg.bisection.prover}, l.~\ref{line:comm});
so that they can update the current node that will be queried in the next step of the bisection game
(Alg.~\ref{alg.bisection.prover}, ll.~\ref{line:descend1} and~\ref{line:descend2}).

Upon reaching a leaf at some index $j$,
the verifier asks both provers to reveal the alleged committees
$S^j$ and $S^{*,j}$ at the pre-image of the respective leaves.
If $j=1$, he inspects whether $S^j$ or $S^{*,j}$ matches $S^0$.
The prover whose alleged first committee is not equal to $S^0$ loses the game.

If $j>1$, the verifier also requests from the provers
(i) the committees at the $(j-1)^\text{th}$ leaves,
(ii) their Merkle proofs with respect to $\mroot$ and $\mroot^*$,
and (iii) the handover proofs $\Sigma^{j}$ and $\Sigma^{*,j}$.
The honest prover responds with (i) $S^{j-1}$ assigned to epoch
$j - 1$, (ii) its Merkle proof with respect to $\mroot$, and
(iii) its own view of the handover proof $\Sigma^j$ (which might
be different from other provers).
Upon checking the Merkle proofs,
the verifier is now convinced that
the committees $S^{j-1}$ and $S^{*,j-1}$ revealed by $P$ and $P^*$
are the same, since their hashes match.
The verifier subsequently
checks if $\Sigma^{j}$ contains more than $\frac{m}{2}$ signatures by the
committee members in $S^{j-1}$ on $(j, S^{j})$,
and similarly for $P^*$.

The prover that fails any of checks by the verifier
loses the bisection game.
If one prover loses the game, and the other one does not fail any checks,
the standing prover is designated the winner.
If neither prover
fails any of the checks, then the verifier concludes that there are over
$\frac{m}{2}$ committee members in $S^{j-1}$ that signed different future sync committees
(\ie, signed both $(j,S^j)$ and $(j,S^{*,j})$, where $(j,S^j) \neq (j,S^{*,j})$).
This implies $S^{j-1}$ is not the correct sync committee assigned to epoch $j-1$,
and both provers are adversarial.
In this case, both provers lose the bisection game.
In any case, at most one prover can win the bisection game.

\begin{algorithm}[tb]
    \caption{Run by the verifier to identify the first different peak in the MMRs of the two provers.
        Here, $\hon{\mroot_{1, \ldots, n}}$ and $\adv{\mroot^*_{1, \ldots, n}}$
        denote the peaks of the honest and adversarial provers respectively.}
    \label{alg.peaks.vs.peaks}
    \begin{algorithmic}[1]\small
        \Function{\sc BisectionGame}{$\hon{P}, \adv{P^*}$}
            \State\Receive{\hon{\mroot_{1, \ldots, n}}}{\hon{P}}
            \State\Receive{\adv{\mroot^*_{1, \ldots, n}}}{\adv{P^*}}
            \For{$i = 1$ to $n$}
                \If{$\hon{\mroot_i} \neq \adv{\mroot^*_i}$}
                    \State\Let{\ell}{\text{size of the $i^\text{th}$ Merkle Tree}}
                    \State\Return{$\textsc{FindDisagreement}(\hon{P}, \hon{\mroot_i}, \adv{P^*}, \adv{\mroot_i^*}, \ell)$}
                \EndIf
            \EndFor
        \EndFunction
    \end{algorithmic}
\end{algorithm}

\myparagraph{Bisection games on Merkle mountain ranges.}
When the number of epochs $N$ is not a power of two, the verifier first obtains the binary
decomposition $\sum_{i=1}^n 2^{q_i} = N$, where $q_1>\ldots>q_n$.
Then, for each prover $P$, he checks if there are $n$ peaks returned.
For two provers $P$ and $P^*$ that have $n$ peaks but returned different commitments,
the verifier compares the peaks $\mroot_i$ of $P$ with $\mroot^*_i$ of $P^*$,
and identifies the first different peak (Alg.~\ref{alg.peaks.vs.peaks}).
It then plays the bisection game as described above on the identified Merkle trees.
The only difference with the game above is that if the disagreement is on the first leaf $j$ of a later tree,
then the Merkle proof for the previous leaf $j-1$ is shown with respect to the peak of the previous tree.

\myparagraph{Prover complexity.}
Given all past sync committees, the prover constructs the MMR in linear time. The MMR is updated in an 
online fashion as time evolves. Every time a new sync committee appears, it is appended to the tree in $\log N$
time. The space required to store the MMR is linear. 

\begin{algorithm}[tb]
    \caption{Tournament conducted by the verifier among provers $\mathcal{P}$
        to identify the 
        state commitment $\stc$.
        The verifier uses
        $\textsc{BisectionGame}$ 
        (\cf Fig.~\ref{fig.bisection-game}, Algs.~\ref{alg.bisection.verifier} and~\ref{alg.peaks.vs.peaks})
        between two provers and deduce at most \emph{one} winner.
        Here, $\mathrm{pop}$ removes and returns an arbitrary element of a set.}
    \label{alg.tournament}
    \begin{algorithmic}[1]\small
        \Function{\sc Tournament}{$\mathcal{P}$}
            \State\Let{P}{\mathrm{pop}(\mathcal{P})};
            \Let{\mathrm{good}}{\{P\}};
            \Let{\stc}{P.\stc}
            \For{$P \in \mathcal{P}$}
                \If{$\stc = P.\stc$}
                    \State\Let{\textrm{good}}{\mathrm{good} \cup \{P\}}
                    \Continue
                \EndIf
                \For{$P^* \in \mathrm{good}$}   \label{line:sample}
                    \If{$\textsc{BisectionGame}(P, P^*) = P$}   \label{line:game}
                        \State\Let{\textrm{good}}{\{P\}};
                        \Let{\stc}{P.\stc}
                        \Break
                    \EndIf
                \EndFor
            \EndFor
            \State\Return{$\stc$}
        \EndFunction
    \end{algorithmic}
\end{algorithm}

\myparagraph{Tournament.}
When there are multiple provers,
the verifier interacts with them sequentially in pairs, in a tournament fashion.
It begins by choosing two provers $P_1$ and $P_2$ with different state commitments
from the set $\mathcal{P}$ (Alg.~\ref{alg.tournament}, l.~\ref{line:sample}).
The verifier then \emph{pits one against the other},
by facilitating a bisection game between $P_1$ and $P_2$,
and decides which of the two provers loses (Alg.~\ref{alg.tournament}, l.~\ref{line:game}).
(There can be at most one winner at any bisection game).
He then eliminates the loser from the tournament,
and chooses a new prover with a different state commitment than the winner's
commitment from the set $\mathcal{P}$ to compete against the winner.
In the event that both provers lose, the verifier eliminates both provers,
and continues the tournament with the remaining ones by sampling two new provers with different
state commitments.
This process continues until all provers left have the same state commitment.
This commitment is adopted as the correct one.
A tournament started with $q$ provers terminates after $O(q)$ bisection games, since at least one prover is eliminated
at the end of each game.
\ifshort
In the full version of the paper~\cite{popos-eprint},
\else
In App.~\ref{sec.proofs}, 
\fi
we prove the security of the tournament by
showing that an honest prover never loses the bisection game and
an adversarial prover loses against an honest one.

\begin{figure}
    \centering
    \includegraphics[width=0.8\linewidth]{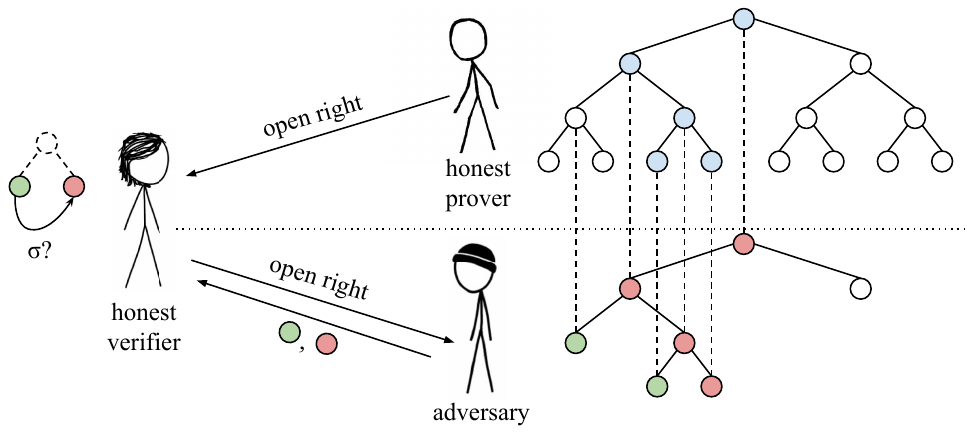}%
    \caption{%
        Honest and adversarial prover in the PoPoS bisection game (\cf Algs.~\ref{alg.bisection.verifier}, \ref{alg.bisection.prover} and~\ref{alg.peaks.vs.peaks}).
        The verifier iteratively requests openings of tree nodes from both provers, until the first
        point of disagreement is discovered.%
    }
    \label{fig.bisection-game}
\end{figure}

\myparagraph{Past and future.}
Now that the verifier obtained the state commitment signed for the most recent epoch,
and confirmed its veracity, the task that remains is to discern facts about
the system's state and its history.
To perform queries about the current state, such as determining how much balance
one owns, the verifier simply asks for Merkle inclusion proofs into the proven
state commitment.

One drawback of our protocol is that the state commitment
received by the verifier is the commitment at the \emph{beginning} of the current
epoch, and therefore may be somewhat stale. In order to synchronize with the latest
state within the epoch, the verifier must function as a full node for the small
duration of an epoch. This functionality does not harm succinctness, since epochs
have a fixed, constant duration. For example, in the case of a longest-chain blockchain,
the protocol works as follows. In addition to signing the state commitment, the sync
committee also signs the first stable block header of its respective epoch. The block header
is verified by the verifier in a similar fashion that he verified the state commitment.
Subsequently, the block header can be used as a \emph{neon genesis} block.
The verifier treats the block as a replacement for the genesis block and bootstraps
from there\footnote{While bootstrapping, the verifier can update the state commitment by applying the transactions
    within the later blocks on top of the state commitment from the neon genesis block via the
    function $\left<\transition\right>$.}.

One aspect of wallets that we have not touched upon concerns the retrieval
and verification of historical transactions. Consider a client that wishes to verify 
the inclusion of a particular historical transaction $\tx$ in the chain. 
Let's assume that $\tx$ is included in a block $B$ of epoch $j$.
This can be checked as follows. 
The verifier, as before, identifies the root of the correct
handover tree. The verifier next asks the prover to provide him with the 
sync committee of epoch $j + 1$, with the corresponding inclusion proof,
as well as the first stable block header $B'$ of that epoch signed by the committee.
Subsequently, he requests the short
blockchain that connects $B$ to $B'$. As blockchains are hash chains,
this inclusion cannot be faked by an adversary.

\section{Proof-of-Stake Ethereum Light Clients}
\label{sec.pos-eth}

The bisection games presented in Sec.~\ref{sec.bisection} can be applied to a variety of PoS consensus protocols to efficiently catch up with
current consensus decisions.
In this section we present an instantiation for Ethereum.
We also detail how to utilize the latest epoch committee to build a full-featured Ethereum JSON-RPC. This allows for existing wallets such as MetaMask to use our construction without making any changes. Our implementation can be a drop-in replacement to obtain better decentralization and performance.

Our PoPoS protocol for Ethereum does not require any changes to the consensus layer, as Ethereum already provisions for sync committees in the way we
introduced in Sec.~\ref{sec.sequences}.

\subsection{Sync Committee Essentials}
\label{sec:poseth-synccommittee}

Sync committees of Ethereum contain $m = 512$ validators, sampled uniformly at random from the validator set, in proportion to their stake distribution. Every sync committee is selected for the duration of a so-called \emph{sync committee period}~\cite{minimal_light_client} (which we called \emph{epoch} in our generic construction).
Each period lasts $256$ Ethereum epochs (these are different from our epochs), approximately $27$ hours.
Ethereum epochs are further divided into \emph{slots}, during which a new block is proposed by one validator and signed by the subset of validators assigned to the slot.
At each slot, each sync committee member of the corresponding period signs the block at the tip of the chain (called the \emph{beacon chain}~\cite{minimal_light_client}) according to its view.
The proposer of the next slot aggregates and includes within its proposal the aggregate sync committee signature on the parent block.
The sync committees are determined one period in advance, and the committee for each period is contained in the block headers of the previous period.
Each block also contains a commitment to the header of the last finalized block that lies on its prefix.

\subsection{Linear-Complexity Light Client}
\label{sec:poseth-lightclient}

Light clients use the sync committee signatures to detect the latest beacon chain block finalized by the Casper FFG finality gadget~\cite{casper,gasper}.
At any round, the view of a light client consists of a $\mathsf{finalized\_header}$, the current sync committee and the next committee.
The client updates its view upon receiving a $\mathsf{LightClientUpdate}$ object (update for short), that contains (i) an $\mathsf{attested\_header}$ signed by the sync committee,
(ii) the corresponding aggregate BLS signature, (iii) the slot at which the aggregate signature was created, (iv) the next sync committee as stated in the $\mathsf{attested\_header}$,
and (v) a $\mathsf{finalized\_header}$ (called the new finalized header for clarity) to replace the one held by the client.

To validate an update, the client first checks if the aggregate signature is from a slot larger than the $\mathsf{finalized\_header}$ in its view, and if this slot is within the current or the next period.
(Updates with signatures from sync committees that are more than one period in the future are rejected.)
It then verifies the inclusion of the new finalized header and the next sync committee provided by the update with respect to the state of the $\mathsf{attested\_header}$ through Merkle inclusion proofs.
Finally, it verifies the aggregate signature on the $\mathsf{attested\_header}$ by the committee of the corresponding period.
Since the signatures are either from the current period or the next one, the client knows the respective committee.

After validating the update, the client replaces its $\mathsf{finalized\_header}$ with the new one, if the $\mathsf{attested\_header}$
was signed by over $2/3$ of the corresponding sync committee.
If this header is from a higher period, the client also updates its view of the sync committees.
Namely, the old next sync committee becomes the new current committee, and the next sync committee included in the $\mathsf{attested\_header}$ is adopted as the new next sync committee.

\subsection{Logarithmic Bootstrapping from Bisection Games}
\label{sec:poseth-bisectiongames}

The construction above requires a bootstrapping light client to download at least one update per period, imposing a linear communication complexity in the life time of the chain.
To reduce the communication load and complexity, the \optclient and superlight client
constructions introduced in Secs.~\ref{sec.sequences} and~\ref{sec.bisection} can be applied to Ethereum.

A bootstrapping superlight client first connects to a few provers, and asks for the Merkle roots of the handover trees (\cf Sec.~\ref{sec.bisection}). The leaf of the handover tree at position $j$ consist of all the public keys of the sync committee of period $j$ concatenated with the period index $j$.
If all the roots are the same, then the client accepts the sync committee at the last leaf as the most recent committee.
If the roots are different, the client facilitates bisection games among conflicting provers.
Upon identifying the first point of disagreement between two trees (\eg, some leaf $j$), the client asks each prover to provide a $\mathsf{LightClientUpdate}$ object to justify the handover from the committee $S^{j-1}$ to $S^j$.
For this purpose, each prover has to provide a valid update that includes (i) an aggregate signature by $2/3$ of the set $S^{j - 1}$ on an $\mathsf{attested\_header}$, and (ii) the set $S^{j}$ as the next sync committee within the $\mathsf{attested\_header}$.
Upon identifying the honest prover, and the correct latest sync committee, the client can ask the honest prover about the lastest update signed by the latest sync committe and containing the tip of the chain.

\subsection{Superlight Client Architecture}
\label{sec:poseth-slc-architecture}

\begin{figure}[tb]
    \centering
    \begin{tikzpicture}[y=2.5cm,x=5cm]
        \small

        \node [color=black!50] at (0,0.5) {\textsc{Server}};
        \node [color=black!50] at (1,0.5) {\textsc{Client}};
        \draw [dashed,draw=black!40] (0.5,-1.5) -- (0.5,0.65);

        \node [align=center,draw,minimum width=2.2cm,fill=red!20] (prover) at (0,0) {Superlight\\client prover};
        \node [align=center,draw,minimum width=2.2cm,fill=black!10] (rpc) at (0,-1) {Full node\\Ethereum\\JSON-RPC};

        \node [align=center,draw,minimum width=2.2cm,fill=red!20] (verifier) at (1,0) {Superlight\\client verifier};
        \node [align=center,draw,minimum width=2.2cm,fill=blue!20] (shim) at (1,-1) {Ethereum\\JSON-RPC\\shim/proxy};

        \node [align=center,draw,fill=black!10,rotate=90] (wallet) at (1.5,-1) {Wallet};
        \node [align=center,rotate=90] (ethereum) at (-0.5,-1) {Ethereum};

        \draw [Latex-Latex] (ethereum) -- (rpc) node [midway,above,rotate=90,anchor=west] {P2P network};
        \draw [Latex-Latex] (shim) -- (wallet) node [midway,above,rotate=90,anchor=west] {JSON-RPC};

        \draw [Latex-Latex,very thick] (prover) -- (verifier) node [midway,above,align=center] {Bisection\\games};
        \draw [Latex-] ([yshift=5pt] rpc.east) -- ([yshift=5pt] shim.west) node [midway,above,align=center] {New txs.};
        \draw [Latex-Latex] ([yshift=-5pt] rpc.east) -- ([yshift=-5pt] shim.west) node [midway,below,align=center] {getProof};
        \draw [-Latex] (rpc) -- (prover) node [midway,right,xshift=-24pt] {Sync info.};
        \draw [-Latex] (verifier) -- (shim) node [midway,left,xshift=16pt] {Consensus tip};

    \end{tikzpicture}%
    \caption{%
        Ethereum superlight client architecture:
        On server side, an Ethereum full node feeds sync information to a bisection game prover sidecar. On client side, a bisection game verifier feeds the consensus tip into an Ethereum JSON-RPC shim/proxy, which forwards transactions coming from the
        wallet to the Ethereum full node, and resolves state queries with reference to the established consensus tip using Ethereum's getProof RPC endpoint.%
    }
    \label{fig:architecture}
\end{figure}
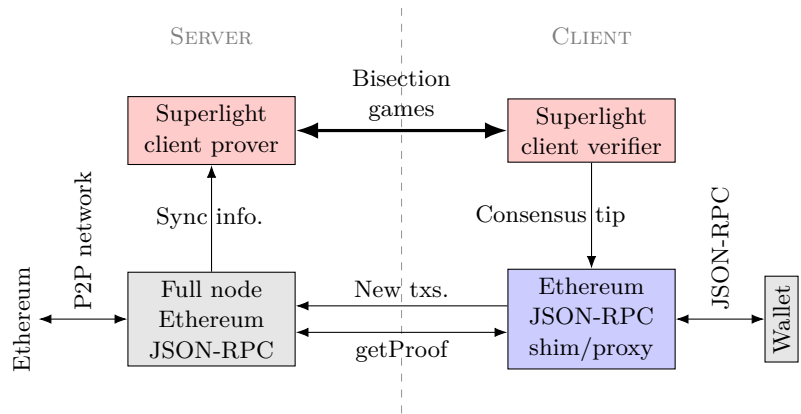

On the completion of bootstrapping, the client has identified the latest beacon chain blockheader. The blockheader contains the commitment to the state of the Ethereum universe that results from executing all transactions since genesis up to and including the present block. Furthermore, this commitment gets verified as part of consensus. The client can perform query to the fullnode about the state of Ethereum. The result of the query can be then verified against the state commitment using Merkle inclusion proofs. This allows for the client to access the state of the Ethereum universe in a trust-minimizing way.

Fig.~\ref{fig:architecture} depicts the resulting architecture of the superlight client. In today's Ethereum, a user's wallet typically speaks to Ethereum JSON-RPC endpoints provided by either a centralized infrastructure provider such as Infura or by a (trusted) Ethereum full node (could be self-hosted). Instead, the centerpiece in a superlight client is a shim that provides RPC endpoints to the wallet, but where new transactions and queries to the Ethereum state are proxied to upstream full nodes, and query responses are verified w.r.t. a given commitment to the Ethereum state. This commitment is produced using two sidecar processes, which implement the prover and verifier of the bisection game. For this purpose, the server-side sidecar obtains the latest sync information from a full node, using what is commonly called `libp2p API'. The client-side sidecar feeds the block header at the consensus tip into the shim.

\section{Experiments}
\label{sec.experiments}

To assess the different bootstrapping mechanisms
for Ethereum
(\expLC = \expLClegend;
\expOLC = \expOLClegend;
\expSLC = \expSLClegend),
we implemented
them
in $\approx2000$ lines of TypeScript code
(source code available on Github\footnote{%
    \ifanonymous{%
        The superlight client prototype,
        optimistic light client,
        and RPC shim
        are at
        \url{https://anonymous.4open.science/r/poc-superlight-client-56E1/},
        \url{https://anonymous.4open.science/r/kevlar-9A68/},
        and
        \url{https://anonymous.4open.science/r/patronum-5187/},
        respectively.
        \textcolor{red}{Links to the productized code bases are omitted for author anonymity.}%
    }%
    \else{%
        The superlight client prototype is at \url{https://github.com/lightclients/poc-superlight-client}.
        The optimistic light client implementation
        is at \url{https://github.com/lightclients/kevlar} and \url{https://kevlar.sh/}.
        The RPC shim is at \url{https://github.com/lightclients/patronum}.%
    }%
    \fi%
}).
We demonstrate an improvement
of \expSLClegend over \expLClegend of
$9\times$ in time-to-completion,
$180\times$ in communication bandwidth,
and
$30\times$ in energy consumption,
when bootstrapping after $10$ years of consensus execution.
\expSLClegend improves over \expOLClegend
by $3\times$ in communication bandwidth
in this setting.

\subsection{Setup}
\label{sec:experiments-setup}

Our experimental scenario includes seven malicious provers, one honest prover, and a verifier. All provers run in different Heroku `\texttt{performance-m}' instances located in the `\texttt{us}' region. The verifier runs on an Amazon EC2 `\texttt{m5.large}' instance located in `\texttt{us-west-2}'. The provers' Internet access is not restricted beyond the hosting provider's limits. The verifier's down- and upload bandwidth is artificially rate-limited to 100\,Mbit/s and 10\,Mbit/s, respectively, using `\texttt{tc}'. We monitor to rule out spillover from RAM into swap space.

In preprocessing, we create eight valid traces of the sync committee protocol for an execution horizon of $30$ years. For this purpose, we create $512$ cryptographic identities per simulated day, as well as the aggregate signatures for handover from one day's sync committee to the next day's.
In some experiments, we vary how much simulated time has passed since genesis, and for this purpose truncate the execution traces accordingly. One of the execution traces is used by the honest prover and understood to be the true honest execution. Adversarial provers each pick a random point in time, and splice the honest execution trace up to that point together with one of the other execution traces for the remaining execution time, \emph{without regenerating handover signatures}, so that the resulting execution trace used by adversarial provers has invalid handover at the point of splicing.
We also vary the internal parameters of the (super-)light client protocols (\ie, batch size $b$ of \expLClegend and \expOLClegend, Merkle tree degree $d$ of \expSLClegend).

\subsection{Time-To-Completion \& Total Verifier Communication}
\label{sec:experiments-tradeoff-delay-bandwidth}

\pgfplotsset{
    myexp1plt/.style={
            nodes near coords,
            point meta=explicit symbolic,
            visualization depends on={value \thisrow{style}\as\mystyle},
            every node near coord/.append style={font=\footnotesize,\mystyle},
        },
}

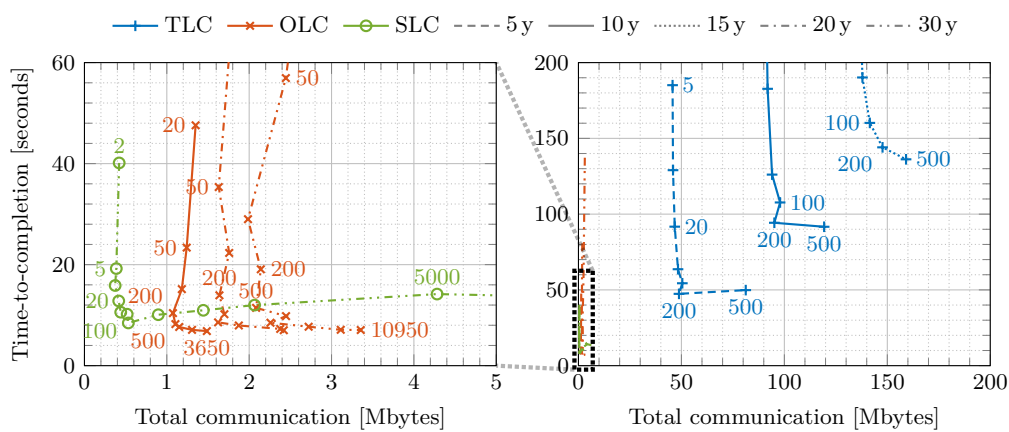
\begin{figure}[tb]
    \centering
    \begin{tikzpicture}
        \begin{axis}[
                mysimpleplot,
                xlabel={Total communication [Mbytes]},
                xmin=0, xmax=200,
                ymin=0, ymax=200,
                width=0.5\linewidth,
                height=0.4\linewidth,
                legend columns=8,
                legend style={
                        at={(0.5,1)},
                        yshift=0.3em,
                        anchor=south,
                        draw=none,
                        /tikz/every even column/.append style={
                                column sep=0.3em
                            },
                        cells={
                                align=left
                            },
                        xshift=-3.25cm,
                    },
            ]

            \addlegendimage{myClientLC1};
            \addlegendentry{\expLClegend};
            \addlegendimage{myClientOLC1};
            \addlegendentry{\expOLClegend};
            \addlegendimage{myClientSLC1};
            \addlegendentry{\expSLClegend};

            \addlegendimage{myClientLC2,mark=none,gray};
            \addlegendentry{5\,y};
            \addlegendimage{myClientLC1,mark=none,gray};
            \addlegendentry{10\,y};
            \addlegendimage{myClientLC3,mark=none,gray};
            \addlegendentry{15\,y};
            \addlegendimage{myClientLC4,mark=none,gray};
            \addlegendentry{20\,y};
            \addlegendimage{myClientLC5,mark=none,gray};
            \addlegendentry{30\,y};

            \addplot+ [
                myexp1plt,
                myClientLC2,
                forget plot,
            ] table [
                    x=bandwidth,
                    y=delay,
                    meta=label,
                ] {figures/plots/experiment-1/light-T1850.txt};

            \addplot+ [
                myexp1plt,
                myClientLC1,
                forget plot,
            ] table [
                    x=bandwidth,
                    y=delay,
                    meta=label,
                ] {figures/plots/experiment-1/light-T3650.txt};

            \addplot+ [
                myexp1plt,
                myClientLC3,
                forget plot,
            ] table [
                    x=bandwidth,
                    y=delay,
                    meta=label,
                ] {figures/plots/experiment-1/light-T5475.txt};

            \addplot+ [
                myexp1plt,
                myClientOLC2,
                no marks,
                forget plot,
            ] table [
                    x=bandwidth,
                    y=delay,
                ] {figures/plots/experiment-1/optimisticlight-T3650.txt};

            \addplot+ [
                myexp1plt,
                myClientOLC4,
                no marks,
                forget plot,
            ] table [
                    x=bandwidth,
                    y=delay,
                ] {figures/plots/experiment-1/optimisticlight-T7300.txt};

            \addplot+ [
                myexp1plt,
                myClientOLC5,
                no marks,
                forget plot,
            ] table [
                    x=bandwidth,
                    y=delay,
                ] {figures/plots/experiment-1/optimisticlight-T10950.txt};

            \addplot+ [
                myexp1plt,
                myClientSLC2,
                no marks,
                forget plot,
            ] table [
                    x=bandwidth,
                    y=delay,
                ] {figures/plots/experiment-1/superlight-T3650.txt};

            \addplot+ [
                myexp1plt,
                myClientSLC4,
                no marks,
                forget plot,
            ] table [
                    x=bandwidth,
                    y=delay,
                ] {figures/plots/experiment-1/superlight-T7300.txt};

            \addplot+ [
                myexp1plt,
                myClientSLC5,
                no marks,
                forget plot,
            ] table [
                    x=bandwidth,
                    y=delay,
                ] {figures/plots/experiment-1/superlight-T10950.txt};

            \coordinate (boxA1) at ([xshift=1.5pt,yshift=1.5pt] 5,60);
            \coordinate (boxB1) at ([xshift=-1.5pt,yshift=-1.5pt] 0,0);

        \end{axis}
        \draw [ultra thick, densely dotted] (boxA1) rectangle (boxB1);
        \begin{axis}[
            mysimpleplot,
            xlabel={Total communication [Mbytes]},
            xmin=0, xmax=5,
            ylabel={Time-to-completion [seconds]},
            ymin=0, ymax=60,
            width=0.5\linewidth,
            height=0.4\linewidth,
            xshift=-6.5cm,
            ]

            \addplot+ [
                myexp1plt,
                myClientOLC1,
            ] table [
                    x=bandwidth,
                    y=delay,
                    meta=label,
                ] {figures/plots/experiment-1/optimisticlight-T3650.txt};

            \addplot+ [
                myexp1plt,
                myClientOLC4,
            ] table [
                    x=bandwidth,
                    y=delay,
                    meta=label,
                ] {figures/plots/experiment-1/optimisticlight-T7300.txt};

            \addplot+ [
                myexp1plt,
                myClientOLC5,
            ] table [
                    x=bandwidth,
                    y=delay,
                    meta=label,
                ] {figures/plots/experiment-1/optimisticlight-T10950.txt};

            \addplot+ [
                myexp1plt,
                myClientSLC5,
            ] table [
                    x=bandwidth,
                    y=delay,
                    meta=label,
                ] {figures/plots/experiment-1/superlight-T10950.txt};

            \legend{};

            \coordinate (boxA2) at (5,60);
            \coordinate (boxB2) at (5,0);

        \end{axis}
        \draw [ultra thick, densely dotted, opacity=0.3] (boxA1) -- (boxA2);
        \draw [ultra thick, densely dotted, opacity=0.3] (boxB1) -- (boxB2);
    \end{tikzpicture}%
    \caption{%
        Time-to-completion and total communication
        (averaged over $5$ trials)
        incurred
        by different light clients
        for varying internal parameters (marker labels; \expLClegend/\expOLClegend: batch size $b$,
        \expSLClegend: Merkle tree degree $d$)
        and varying consensus execution horizon.
        For \expSLClegend, the curves for the considered execution horizons are virtually identical;
        thus, only the curve for $30$ years (most challenging scenario) is shown.
        Pareto-optimal tradeoffs are
        at `tip' of resulting
        `L-shape':
        for $10$ years execution,
        at
        $b\approx200$ (\expLClegend), $b\approx500$ (\expOLClegend), and $d\approx100$ (\expSLClegend),
        respectively.
        \expOLClegend and \expSLClegend vastly outperform
        \expLClegend,
        \eg,
        for $10$ years execution:
        $9\times$ in time-to-completion,
        $180\times$ in bandwidth.
        In this setting, \expSLClegend has similar time-to-completion as \expOLClegend,
        and $3\times$ lower communication.
    }
    \label{fig:experiment-experiment-1}
\end{figure}

The average
time-to-completion (TTC) and total communication bandwidth (TCB)
required
by the different light client constructions
per bootstrapping occurrence
is plotted in Fig.~\ref{fig:experiment-experiment-1}
for varying internal parameters
(batch sizes $b$ for \expLClegend and \expOLClegend; Merkle tree degrees $d$ for \expSLClegend)
and varying execution horizons (from $5$ to $30$ years).
Pareto-optimal
TTC and TCB
are achieved for $b$ and $d$
resulting
`at the tip' of the
`L-shaped' plot.
For instance,
for $10$ years execution,
\expLClegend, \expOLClegend and \expSLClegend
achieve Pareto-optimal TTC/TCB for
$b\approx200$, $b\approx500$, and $d\approx100$,
respectively.
Evidently, across a wide parameter range,
\expOLClegend and \expSLClegend vastly outperform
\expLClegend in both metrics; \eg,
for $10$ years execution and Pareto-optimal parameters,
$9\times$ in TTC, and
$180\times$ in TCB.
In this setting, \expSLClegend has similar TTC as \expOLClegend,
and $3\times$ lower TCB ($5\times$ lower TCB for $30$ years).
For a closed-form expression describing the trade-off between
latency and bandwidth, and the optimal choice of tree degree
see Tas et al.~\cite{lazylight}.

\begin{figure}[tb]
    \centering
    \begin{tikzpicture}
        \begin{axis}[
            mysimpleplot,
            xlabel={Execution horizon [years]},
            xmode=log,
            xmin=0, xmax=15000,
            xtick={684,1368,2737,5475,10950},
            xticklabels={1.875, 3.75, 7.5, 15, 30},
            ylabel={Time-to-completion [seconds]},
            ymin=0, ymax=150,
            width=0.7\linewidth,
            height=0.35\linewidth,
            ]

            \addplot+ [
                nodes near coords,
                point meta=explicit symbolic,
                visualization depends on={value \thisrow{style}\as\mystyle},
                every node near coord/.append style={font=\footnotesize,\mystyle},
                myClientOLC1,
            ] table [
                    x=execution_horizon,
                    y=delay,
                    meta=label,
                ] {figures/plots/experiments-23/optimisticlight.txt};
            \addlegendentry{\expOLClegend ($b=20$)};

            \addplot+ [
                nodes near coords,
                point meta=explicit symbolic,
                visualization depends on={value \thisrow{style}\as\mystyle},
                every node near coord/.append style={font=\footnotesize,\mystyle},
                myClientSLC1,
            ] table [
                    x=execution_horizon,
                    y=delay,
                    meta=label,
                ] {figures/plots/experiments-23/superlight.txt};
            \addlegendentry{\expSLClegend ($d=2$)};

        \end{axis}
    \end{tikzpicture}%
    \vspace{-0.75em}%
    \caption{%
        Time-to-completion
        (averaged over $5$ trials)
        of \expOLClegend/\expSLClegend
        increase linearly/logarithmically
        with the execution horizon, respectively.
    }
    \label{fig:experiment-experiments-23}
\end{figure}
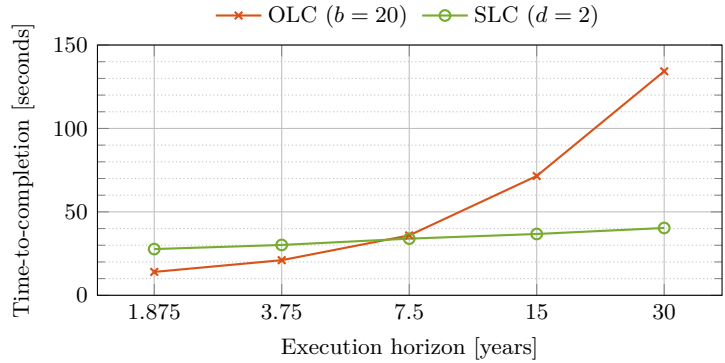

The fact that both \expLClegend and \expOLClegend
have TCB linear in the execution horizon,
is readily apparent from
Fig.~\ref{fig:experiment-experiment-1}.
The linear TTC is visible for \expLClegend,
but not very pronounced for \expOLClegend,
due to the concretely low proportionality constant.
In comparison, \expSLClegend shows barely any dependence
of TTC or TCB on the execution horizon,
hinting at the (exponentially better) logarithmic dependence.
To contrast the asymptotics, we plot average
TTC as a function of exponentially increasing
execution horizon
in Fig.~\ref{fig:experiment-experiments-23}
for \expOLClegend and \expSLClegend
with internal parameters
$b=20$ and $d=2$, respectively.
Note that these are not Pareto-optimal parameters,
but chosen here for illustration purposes.
Clearly, TTC for \expOLClegend is linear in the
execution horizon (plotted
in Fig.~\ref{fig:experiment-experiments-23}
on an exponential scale),
while for \expSLClegend it is logarithmic.

\subsection{Power \& Energy Consumption}
\label{sec:experiments-energy}

\pgfplotsset{
    mybarchart01/.style={
            enlarge x limits=0.15,
            ybar,
            bar width=6mm,
            xtick=data,
            width=\linewidth,
            height=0.6\linewidth,
            ymin=0,
            grid=both,
            minor y tick num=4,
            minor x tick num=1,
            major grid style={solid,draw=gray!50},
            minor grid style={densely dotted,draw=gray!50},
            label style={font=\small},
            tick label style={font=\small},
        },
}

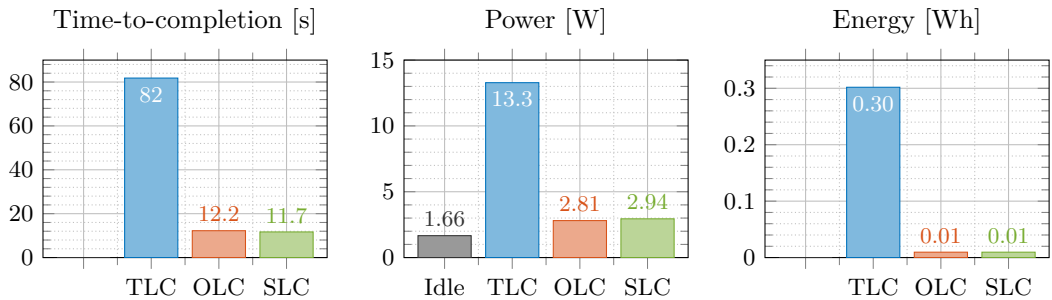
\begin{figure}[tb]
    \centering
    \begin{tikzpicture}
        \begin{axis}[
            mybarchart01,
            title={Time-to-completion [$\mathrm{s}$]},
            xtick={1,2,3,4},
            xticklabel style={align=center},
            xticklabels={{}, {\expLClegend}, {\expOLClegend}, {\expSLClegend}},
            enlarge x limits=0.2,
            bar width=7mm,
            ymin=0.00, ymax=90,
            yticklabel style={
                    /pgf/number format/fixed,
                    /pgf/number format/precision=2,
                },
            scaled y ticks=false,
            width=0.38\linewidth,
            height=0.3\linewidth,
            label style={
                    align=center,
                },
            xshift=-4.75cm,
            ]

            \addplot+ [
                bar shift=0pt,
                color=myClientColorIdle,
                fill=myClientColorIdle!50,
            ] coordinates {
                    (1,0)   %
                };

            \addplot+ [
                bar shift=0pt,
                color=myClientColorLC,
                fill=myClientColorLC!50,
            ] coordinates {
                    (2,81.8)
                };
            \node [below,white,font=\footnotesize] at (axis cs:2,81.8) {$82$};

            \addplot+ [
                bar shift=0pt,
                color=myClientColorOLC,
                fill=myClientColorOLC!50,
            ] coordinates {
                    (3,12.24)
                };
            \node [above,myClientColorOLC,font=\footnotesize] at (axis cs:3,12.24) {$12.2$};

            \addplot+ [
                bar shift=0pt,
                color=myClientColorSLC,
                fill=myClientColorSLC!50,
            ] coordinates {
                    (4,11.68)
                };
            \node [above,myClientColorSLC,font=\footnotesize] at (axis cs:4,11.68) {$11.7$};

        \end{axis}
        \begin{axis}[
                mybarchart01,
                title={Power [$\mathrm{W}$]},
                xtick={1,2,3,4},
                xticklabel style={align=center},
                xticklabels={Idle, {\expLClegend}, {\expOLClegend}, {\expSLClegend}},
                enlarge x limits=0.2,
                bar width=7mm,
                ymin=0.00, ymax=15,
                yticklabel style={
                        /pgf/number format/fixed,
                        /pgf/number format/precision=2
                    },
                scaled y ticks=false,
                width=0.38\linewidth,
                height=0.3\linewidth,
                label style={
                        align=center,
                    },
            ]

            \addplot+ [
                bar shift=0pt,
                color=myClientColorIdle,
                fill=myClientColorIdle!50,
            ] coordinates {
                    (1,1.6632000000000033)
                };
            \node [above,myClientColorIdle,font=\footnotesize] at (axis cs:1,1.6632000000000033) {$1.66$};

            \addplot+ [
                bar shift=0pt,
                color=myClientColorLC,
                fill=myClientColorLC!50,
            ] coordinates {
                    (2,13.283911980440159)
                };
            \node [below,white,font=\footnotesize] at (axis cs:2,13.283911980440159) {$13.3$};

            \addplot+ [
                bar shift=0pt,
                color=myClientColorOLC,
                fill=myClientColorOLC!50,
            ] coordinates {
                    (3,2.8082352941175808)
                };
            \node [above,myClientColorOLC,font=\footnotesize] at (axis cs:3,2.8082352941175808) {$2.81$};

            \addplot+ [
                bar shift=0pt,
                color=myClientColorSLC,
                fill=myClientColorSLC!50,
            ] coordinates {
                    (4,2.942876712328698)
                };
            \node [above,myClientColorSLC,font=\footnotesize] at (axis cs:4,2.942876712328698) {$2.94$};

        \end{axis}
        \begin{axis}[
                mybarchart01,
                title={Energy [$\mathrm{Wh}$]},
                xtick={1,2,3,4},
                xticklabel style={align=center},
                xticklabels={{}, {\expLClegend}, {\expOLClegend}, {\expSLClegend}},
                enlarge x limits=0.2,
                bar width=7mm,
                ymin=0.00, ymax=0.35,
                yticklabel style={
                        /pgf/number format/fixed,
                        /pgf/number format/precision=2
                    },
                scaled y ticks=false,
                width=0.38\linewidth,
                height=0.3\linewidth,
                label style={
                        align=center,
                    },
                xshift=4.75cm,
            ]

            \addplot+ [
                bar shift=0pt,
                color=myClientColorIdle,
                fill=myClientColorIdle!50,
            ] coordinates {
                    (1,0)   %
                };

            \addplot+ [
                bar shift=0pt,
                color=myClientColorLC,
                fill=myClientColorLC!50,
            ] coordinates {
                    (2,0.3018400000000014)
                };
            \node [below,white,font=\footnotesize] at (axis cs:2,0.3018400000000014) {$0.30$};

            \addplot+ [
                bar shift=0pt,
                color=myClientColorOLC,
                fill=myClientColorOLC!50,
            ] coordinates {
                    (3,0.009547999999999774)
                };
            \node [above,myClientColorOLC,font=\footnotesize] at (axis cs:3,0.009547999999999774) {$0.01$};

            \addplot+ [
                bar shift=0pt,
                color=myClientColorSLC,
                fill=myClientColorSLC!50,
            ] coordinates {
                    (4,0.009547999999999774)
                };
            \node [above,myClientColorSLC,font=\footnotesize] at (axis cs:4,0.009547999999999774) {$0.01$};

        \end{axis}
    \end{tikzpicture}%
    \vspace{-0.75em}%
    \caption{%
        Energy
        required to bootstrap after
        $10$ years of consensus execution
        using different light client constructions
        (averaged over $5$ trials for \expLClegend,
        $25$ trials for \expOLClegend and \expSLClegend;
        internal parameters
        $b=200$, $b=500$, $d=100$, respectively);
        also disaggregated into power consumption
        and time-to-completion.
        Energy required by \expOLClegend/\expSLClegend
        is $30\times$ lower than \expLClegend.
        Contributions $\approx4\times$
        and $\approx7\times$ can be attributed
        to lower power consumption
        and lower time-to-completion,
        respectively.
    }
    \label{fig:experiment-energypower-1}
\end{figure}

A key motivation for superlight clients
is their application on resource-constrained
platforms such as browsers or mobile phones.
In this context, computational efficiency,
and as a proxy energy efficiency, is an important metric.
We ran the light clients
on a battery-powered
System76 Lemur Pro (`\texttt{lemp10}') laptop
with Pop!\_OS 22.04 LTS, and recorded
the decaying battery level using `\texttt{upower}'
(screen off, no other programs running, no
keyboard/mouse input, WiFi connectivity; provers still on Heroku instances).
From the energy consumption and wallclock time
we calculated the average power consumption.
As internal parameters for \expLClegend,
\expOLClegend, and \expSLClegend,
we chose
$b=200$, $b=500$, and $d=100$,
respectively (\cf Pareto-optimal
parameters in Fig.~\ref{fig:experiment-experiment-1}).
The energy required
to bootstrap
$10$ years of consensus execution,
averaged over
$5$ trials for \expLClegend,
and
$25$ trials for \expOLClegend and \expSLClegend,
is plotted in Fig.~\ref{fig:experiment-energypower-1}.
We disaggregate the energy consumption
into power consumption and TTC for each light client,
and also record the power consumption of the machine
in idle.
(Note, discrepancies in Figs.~\ref{fig:experiment-experiment-1} and~\ref{fig:experiment-energypower-1} are due to the light clients running on Amazon EC2 vs. a laptop.)

\expOLClegend and \expSLClegend have comparable
TTC and power consumption, resulting in comparable
energy consumption per bootstrap occurrence.
The energy required by \expOLClegend and \expSLClegend
is $30\times$ lower than the energy required by
\expLClegend per bootstrap occurrence (right panel in Fig.~\ref{fig:experiment-energypower-1}).
This can be attributed to a $\approx4\times$ lower power consumption (middle panel in Fig.~\ref{fig:experiment-energypower-1})
together with a $\approx7\times$ lower TTC (left panel in Fig.~\ref{fig:experiment-energypower-1}).
The considerably lower energy/power consumption
of \expOLClegend/\expSLClegend compared to
\expLClegend
is due to the lower number of signature verifications
(and thus lower computational burden).
Note that a sizeable fraction of \expOLClegend's/\expSLClegend's power consumption
can be attributed to system idle (middle panel in Fig.~\ref{fig:experiment-energypower-1}).
When comparing light clients in terms of
\emph{excess} energy consumption (\ie, subtracting
idle consumption) per bootstrapping,
then \expOLClegend and \expSLClegend
improve over \expSLClegend by $64\times$.

\section{Analysis}
\label{sec.analysis}

The theorems for succinctness and security of the PoPoS protocol are provided below.
Proofs are in 
\ifshort
the full version of this paper~\cite{popos-eprint}.
\else
App.~\ref{sec.proofs}.
\fi
Security consists of two components: completeness and soundness.

\begin{restatable}[Succinctness]{theorem}{restateSuccinctness}
    \label{thm:succinctness}
    Consider a verifier that invokes a bisection game at round $r$ between two provers
    that provided different handover tree roots.
    Then, the game ends in $O(\log(r))$ steps of interactivity and
    has a total communication complexity of $O(\log(r))$.
\end{restatable}

\begin{restatable}[Completeness]{theorem}{restateCompleteness}
    \label{thm:completeness}
    Consider a verifier that invokes a bisection game at round $r$ between two provers
    that provided different handover tree roots.
    Suppose one of the provers is honest.
    Then, the honest prover wins the bisection game.
\end{restatable}

\begin{restatable}[Soundness]{theorem}{restateSoundness}
    \label{thm:soundness}
    Let $H^s$ be a collision resistant hash function.
    Consider a verifier that invokes a bisection game
    executed
    at round $r$
    of a secure underlying PoS protocol
    between two provers
    that provided different handover tree roots.
    Suppose one of the provers is honest, and the signature scheme satisfies existential unforgeability.
    Then, for all PPT adversarial provers $\mathcal{A}$, the prover $\mathcal{A}$ loses the bisection game
    against the honest prover with overwhelming probability in $\lambda$.
\end{restatable}

\begin{restatable}[Tournament Runtime]{theorem}{restateTournamentRuntime}
    \label{thm:tournament-runtime}
    Consider a tournament ran at round $r$ with $|\mathcal{P}|$ provers
    one of which is honest.
    The tournament ends in $O(|\mathcal{P}|\log(r))$ steps of interactivity,
    and has total communication complexity $O(|\mathcal{P}|\log(r))$.
\end{restatable}

\begin{restatable}[Security]{theorem}{restateSecurity}
    \label{thm:security}
    Let $H^s$ be a collision resistant hash function.
    Consider a tournament executed between an honest verifier
    and
    $|\mathcal{P}|$ provers
    at round $r$.
    Suppose one of the provers is honest, the signature scheme satisfies existential unforgeability,
    and the PoS protocol is secure.
    Then, for all PPT adversaries $\mathcal{A}$,
    the state commitment obtained by the verifier at the end of the tournament satisfies
    state security with overwhelming probability in $\lambda$.
\end{restatable}

\clearpage

\bibliography{references}

\begin{thebibliography}{10}

\bibitem{sidechains}
Adam Back, Matt Corallo, Luke Dashjr, Mark Friedenbach, Gregory Maxwell, Andrew
  Miller, Andrew Poelstra, Jorge Tim{\'o}n, and Pieter Wuille.
\newblock Enabling blockchain innovations with pegged sidechains.
\newblock 2014.
\newblock \url{https://blockstream.com/sidechains.pdf}.

\bibitem{genesis}
Christian Badertscher, Peter Gazi, Aggelos Kiayias, Alexander Russell, and
  Vassilis Zikas.
\newblock Ouroboros genesis: Composable proof-of-stake blockchains with dynamic
  availability.
\newblock In {\em {CCS}}, pages 913--930. {ACM}, 2018.

\bibitem{crypto-book-boneh}
Dan Boneh and Victor Shoup.
\newblock {\em {A Graduate Course in Applied Cryptography}}.
\newblock 2020.

\bibitem{coda}
Joseph Bonneau, Izaak Meckler, Vanishree Rao, and Evan Shapiro.
\newblock Coda: Decentralized cryptocurrency at scale.
\newblock Cryptology ePrint Archive, Paper 2020/352, 2020.
\newblock URL: \url{https://eprint.iacr.org/2020/352}.

\bibitem{tendermint-paper}
Ethan Buchman, Jae Kwon, and Zarko Milosevic.
\newblock The latest gossip on bft consensus, 2018.
\newblock \href {https://arxiv.org/abs/1807.04938v3}
  {\path{arXiv:1807.04938v3}}.

\bibitem{flyclient}
Benedikt B{\"{u}}nz, Lucianna Kiffer, Loi Luu, and Mahdi Zamani.
\newblock Flyclient: Super-light clients for cryptocurrencies.
\newblock In {\em {IEEE} Symposium on Security and Privacy}, pages 928--946.
  {IEEE}, 2020.

\bibitem{buterin}
Vitalik Buterin.
\newblock A next-generation smart contract and decentralized application
  platform.
\newblock 2014.

\bibitem{weak-subjectivity}
Vitalik Buterin.
\newblock {Proof of Stake: How I Learned to Love Weak Subjectivity}, Nov 2014.
\newblock URL:
  \url{https://blog.ethereum.org/2014/11/25/proof-stake-learned-love-weak-subjectivity/}.

\bibitem{casper}
Vitalik Buterin and Virgil Griffith.
\newblock Casper the friendly finality gadget, 2017.
\newblock \href {https://arxiv.org/abs/1710.09437v4}
  {\path{arXiv:1710.09437v4}}.

\bibitem{gasper}
Vitalik Buterin, Diego Hernandez, Thor Kamphefner, Khiem Pham, Zhi Qiao, Danny
  Ryan, Juhyeok Sin, Ying Wang, and Yan~X Zhang.
\newblock Combining ghost and casper, 2020.
\newblock \href {https://arxiv.org/abs/2003.03052v3}
  {\path{arXiv:2003.03052v3}}.

\bibitem{practical-delegation}
Ran Canetti, Ben Riva, and Guy~N. Rothblum.
\newblock Practical delegation of computation using multiple servers.
\newblock In {\em {CCS}}, pages 445--454. {ACM}, 2011.

\bibitem{refereed-computation}
Ran Canetti, Ben Riva, and Guy~N. Rothblum.
\newblock Refereed delegation of computation.
\newblock {\em Inf. Comput.}, 226:16--36, 2013.

\bibitem{pbft}
Miguel Castro and Barbara Liskov.
\newblock Practical byzantine fault tolerance.
\newblock In {\em {OSDI}}, pages 173--186. {USENIX} Association, 1999.

\bibitem{mithril}
Pyrros Chaidos and Aggelos Kiayias.
\newblock Mithril: Stake-based threshold multisignatures.
\newblock Cryptology ePrint Archive, Paper 2021/916, 2021.
\newblock URL: \url{https://eprint.iacr.org/2021/916}.

\bibitem{sok-light-clients}
Panagiotis Chatzigiannis, Foteini Baldimtsi, and Konstantinos Chalkias.
\newblock Sok: Blockchain light clients.
\newblock In {\em Financial Cryptography}, volume 13411 of {\em LNCS}, pages
  615--641. Springer, 2022.

\bibitem{algorand}
Jing Chen and Silvio Micali.
\newblock Algorand, 2016.
\newblock \href {https://arxiv.org/abs/1607.01341v9}
  {\path{arXiv:1607.01341v9}}.

\bibitem{metamask-mau-21}
{ConsenSys}.
\newblock {MetaMask Surpasses 10 Million MAUs, Making It The World's Leading
  Non-Custodial Crypto Wallet}, Aug 2021.
\newblock URL:
  \url{https://consensys.net/blog/press-release/metamask-surpasses-10-million-maus-making-it-the-worlds-leading-non-custodial-crypto-wallet/}.

\bibitem{sparse-mt}
Rasmus Dahlberg, Tobias Pulls, and Roel Peeters.
\newblock Efficient sparse merkle trees - caching strategies and secure
  (non-)membership proofs.
\newblock In {\em NordSec}, volume 10014 of {\em LNCS}, pages 199--215, 2016.

\bibitem{snowwhite}
Phil Daian, Rafael Pass, and Elaine Shi.
\newblock Snow white: Robustly reconfigurable consensus and applications to
  provably secure proof of stake.
\newblock Cryptology ePrint Archive, Paper 2016/919, 2016.
\newblock URL: \url{https://eprint.iacr.org/2016/919}.

\bibitem{gas-efficient}
Stelios Daveas, Kostis Karantias, Aggelos Kiayias, and Dionysis Zindros.
\newblock A gas-efficient superlight bitcoin client in solidity.
\newblock In {\em {AFT}}, pages 132--144. {ACM}, 2020.

\bibitem{praos}
Bernardo David, Peter Gazi, Aggelos Kiayias, and Alexander Russell.
\newblock Ouroboros praos: An adaptively-secure, semi-synchronous
  proof-of-stake blockchain.
\newblock In {\em {EUROCRYPT} {(2)}}, volume 10821 of {\em LNCS}, pages 66--98.
  Springer, 2018.

\bibitem{long-range-survey}
Evangelos Deirmentzoglou, Georgios Papakyriakopoulos, and Constantinos
  Patsakis.
\newblock A survey on long-range attacks for proof of stake protocols.
\newblock {\em {IEEE} Access}, 7:28712--28725, 2019.

\bibitem{mmr-grin}
Grin Developers.
\newblock {Merkle Mountain Ranges (MMR)}.
\newblock URL:
  \url{https://docs.grin.mw/wiki/chain-state/merkle-mountain-range/}.

\bibitem{eth-lightclient}
{Ethereum Developers}.
\newblock {Altair Light Client -- Light Client}, 2023.
\newblock URL:
  \url{https://github.com/ethereum/consensus-specs/blob/5c64a2047af9315db4ce3bd0eec0d81194311e46/specs/altair/light-client/light-client.md}.

\bibitem{minimal_light_client}
{Ethereum Developers}.
\newblock {Altair Light Client -- Sync Protocol}, 2023.
\newblock URL:
  \url{https://github.com/ethereum/consensus-specs/blob/e9f1d56807d52aa7425f10160a45cb522345468b/specs/altair/light-client/sync-protocol.md}.

\bibitem{fiatshamir}
Amos Fiat and Adi Shamir.
\newblock {How to prove yourself: Practical solutions to identification and
  signature problems}.
\newblock In {\em Conference on the theory and application of cryptographic
  techniques}, pages 186--194. Springer, 1986.

\bibitem{plumo}
Ariel Gabizon, Kobi Gurkan, Philipp Jovanovic, Georgios Konstantopoulos, Asa
  Oines, Marek Olszewski, Michael Straka, Eran Tromer, and Psi Vesely.
\newblock Plumo: Towards scalable interoperable blockchains using ultra light
  validation systems.
\newblock 2020.
\newblock URL:
  \url{https://docs.zkproof.org/pages/standards/accepted-workshop3/proposal-plumo\_celolightclient.pdf}.

\bibitem{backbone-new}
Juan Garay, Aggelos Kiayias, and Nikos Leonardos.
\newblock The bitcoin backbone protocol: Analysis and applications.
\newblock Cryptology ePrint Archive, Paper 2014/765, 2014.
\newblock URL: \url{https://eprint.iacr.org/2014/765}.

\bibitem{backbone}
Juan~A. Garay, Aggelos Kiayias, and Nikos Leonardos.
\newblock The bitcoin backbone protocol: Analysis and applications.
\newblock In {\em {EUROCRYPT} {(2)}}, volume 9057 of {\em LNCS}, pages
  281--310. Springer, 2015.

\bibitem{varbackbone}
Juan~A. Garay, Aggelos Kiayias, and Nikos Leonardos.
\newblock The bitcoin backbone protocol with chains of variable difficulty.
\newblock In {\em {CRYPTO} {(1)}}, volume 10401 of {\em LNCS}, pages 291--323.
  Springer, 2017.

\bibitem{pos-sidechains}
Peter Gazi, Aggelos Kiayias, and Dionysis Zindros.
\newblock Proof-of-stake sidechains.
\newblock In {\em {IEEE} Symposium on Security and Privacy}, pages 139--156.
  {IEEE}, 2019.

\bibitem{key-evolving}
Gene Itkis and Leonid Reyzin.
\newblock Forward-secure signatures with optimal signing and verifying.
\newblock In {\em {CRYPTO}}, volume 2139 of {\em LNCS}, pages 332--354.
  Springer, 2001.

\bibitem{arbitrum}
Harry~A. Kalodner, Steven Goldfeder, Xiaoqi Chen, S.~Matthew Weinberg, and
  Edward~W. Felten.
\newblock Arbitrum: Scalable, private smart contracts.
\newblock In {\em {USENIX} Security Symposium}, pages 1353--1370. {USENIX}
  Association, 2018.

\bibitem{wallets}
Kostis Karantias.
\newblock Sok: A taxonomy of cryptocurrency wallets.
\newblock Cryptology ePrint Archive, Paper 2020/868, 2020.
\newblock URL: \url{https://eprint.iacr.org/2020/868}.

\bibitem{compactsuperblocks}
Kostis Karantias, Aggelos Kiayias, and Dionysis Zindros.
\newblock Compact storage of superblocks for nipopow applications.
\newblock In {\em {MARBLE}}, pages 77--91. Springer, 2019.

\bibitem{burn}
Kostis Karantias, Aggelos Kiayias, and Dionysis Zindros.
\newblock Proof-of-burn.
\newblock In {\em Financial Cryptography}, volume 12059 of {\em LNCS}, pages
  523--540. Springer, 2020.

\bibitem{katz}
Jonathan Katz and Yehuda Lindell.
\newblock {\em Introduction to Modern Cryptography, Second Edition}.
\newblock {CRC} Press, 2014.

\bibitem{popow}
Aggelos Kiayias, Nikolaos Lamprou, and Aikaterini{-}Panagiota Stouka.
\newblock Proofs of proofs of work with sublinear complexity.
\newblock In {\em Financial Cryptography Workshops}, volume 9604 of {\em LNCS},
  pages 61--78. Springer, 2016.

\bibitem{logspace}
Aggelos Kiayias, Nikos Leonardos, and Dionysis Zindros.
\newblock Mining in logarithmic space.
\newblock In {\em {CCS}}, pages 3487--3501. {ACM}, 2021.

\bibitem{nipopows}
Aggelos Kiayias, Andrew Miller, and Dionysis Zindros.
\newblock Non-interactive proofs of proof-of-work.
\newblock In {\em Financial Cryptography}, volume 12059 of {\em LNCS}, pages
  505--522. Springer, 2020.

\bibitem{velvet-nipopows}
Aggelos Kiayias, Andrianna Polydouri, and Dionysis Zindros.
\newblock The velvet path to superlight blockchain clients.
\newblock In {\em {AFT}}, pages 205--218. {ACM}, 2021.

\bibitem{ouroboros}
Aggelos Kiayias, Alexander Russell, Bernardo David, and Roman Oliynykov.
\newblock Ouroboros: {A} provably secure proof-of-stake blockchain protocol.
\newblock In {\em {CRYPTO} {(1)}}, volume 10401 of {\em LNCS}, pages 357--388.
  Springer, 2017.

\bibitem{pow-sidechains}
Aggelos Kiayias and Dionysis Zindros.
\newblock Proof-of-work sidechains.
\newblock In {\em Financial Cryptography Workshops}, volume 11599 of {\em
  LNCS}, pages 21--34. Springer, 2019.

\bibitem{cosmos-whitepaper}
Jae Kwon and Ethan Buchman.
\newblock A network of distributed ledgers -- cosmos whitepaper.
\newblock URL: \url{https://v1.cosmos.network/resources/whitepaper}.

\bibitem{horizon}
Rongjian Lan, Ganesha Upadhyaya, Stephen Tse, and Mahdi Zamani.
\newblock Horizon: A gas-efficient, trustless bridge for cross-chain
  transactions, 2021.
\newblock \href {https://arxiv.org/abs/2101.06000v1}
  {\path{arXiv:2101.06000v1}}.

\bibitem{merkle}
Ralph~C. Merkle.
\newblock A digital signature based on a conventional encryption function.
\newblock In {\em {CRYPTO}}, volume 293 of {\em LNCS}, pages 369--378.
  Springer, 1987.

\bibitem{pass-asynchronous}
Rafael Pass, Lior Seeman, and Abhi Shelat.
\newblock Analysis of the blockchain protocol in asynchronous networks.
\newblock In {\em {EUROCRYPT} {(2)}}, volume 10211 of {\em LNCS}, pages
  643--673, 2017.

\bibitem{succinct-labs}
{Succinct Labs}.
\newblock {Building the end game of interoperability with zkSNARKs}, 2023.
\newblock URL: \url{https://www.succinct.xyz/}.

\bibitem{lazylight}
Ertem~Nusret Tas, Dionysis Zindros, Lei Yang, and David Tse.
\newblock Light clients for lazy blockchains.
\newblock Cryptology ePrint Archive, Paper 2022/384, 2022.
\newblock URL: \url{https://eprint.iacr.org/2022/384}.

\bibitem{mmr}
Peter Todd.
\newblock Merkle mountain ranges, oct 2012.
\newblock URL:
  \url{https://github.com/opentimestamps/opentimestamps-server/blob/master/doc/merkle-mountain-range.md}.

\bibitem{metamask-mau-22}
Jason Wise.
\newblock {Metamask Statistics 2023: How Many People Use Metamask?}, Mar 2023.
\newblock URL: \url{https://earthweb.com/metamask-statistics/}.

\bibitem{wood}
Gavin Wood.
\newblock Ethereum: A secure decentralised generalised transaction ledger.
\newblock 2014.

\bibitem{zkbridge}
Tiancheng Xie, Jiaheng Zhang, Zerui Cheng, Fan Zhang, Yupeng Zhang, Yongzheng
  Jia, Dan Boneh, and Dawn Song.
\newblock zkbridge: Trustless cross-chain bridges made practical.
\newblock In {\em {CCS}}, pages 3003--3017. {ACM}, 2022.

\bibitem{hotstuff}
Maofan Yin, Dahlia Malkhi, Michael~K. Reiter, Guy Golan{-}Gueta, and Ittai
  Abraham.
\newblock Hotstuff: {BFT} consensus with linearity and responsiveness.
\newblock In {\em {PODC}}, pages 347--356. {ACM}, 2019.

\bibitem{crosschain-sok}
Alexei Zamyatin, Mustafa Al{-}Bassam, Dionysis Zindros, Eleftherios
  Kokoris{-}Kogias, Pedro Moreno{-}Sanchez, Aggelos Kiayias, and William~J.
  Knottenbelt.
\newblock Sok: Communication across distributed ledgers.
\newblock In {\em Financial Cryptography {(2)}}, volume 12675 of {\em LNCS},
  pages 3--36. Springer, 2021.

\bibitem{velvet}
Alexei Zamyatin, Nicholas Stifter, Aljosha Judmayer, Philipp Schindler,
  Edgar~R. Weippl, and William~J. Knottenbelt.
\newblock A wild velvet fork appears! inclusive blockchain protocol changes in
  practice - (short paper).
\newblock In {\em Financial Cryptography Workshops}, volume 10958 of {\em
  LNCS}, pages 31--42. Springer, 2018.

\bibitem{nearbridge}
Maksym Zavershynskyi.
\newblock {ETH-NEAR Rainbow Bridge}, Aug 2020.
\newblock URL: \url{https://near.org/blog/eth-near-rainbow-bridge/}.

\end{thebibliography}

\ifshort
\else
\appendix
\section{Security of Ethereum Light Clients}
\label{sec:sync-committee-security}

The following assumptions ensure the security of the \optclient and superlight client on Ethereum:
\begin{enumerate}
      \item The honest Ethereum validators constitutes at least $\frac{2}{3}+\epsilon$ fraction of the validator set at all times.
      \item The sync committee for each period is sampled uniformly at random from the validator set.
      \item The underlying PoS consensus protocol satisfies security.
      \item The $\mathsf{attested\_header}$ of a beacon block containing a $\mathsf{finalized\_header}$ is signed by a
            sync committee member \emph{only if} the $\mathsf{finalized\_header}$ is the header of a Casper FFG finalized PoS block
            in the view of the sync committee member.
      \item Honest block proposers include the latest Casper FFG finalized block in their view as the $\mathsf{finalized\_header}$
            of their proposal blocks.
\end{enumerate}

The assumptions (a) and (b) ensure that the honest sync committee members constitute a supermajority of the sync
committee at all periods.
Assumption (d) ensures that any header obtained by a light client belongs to a Casper FFG finalized block,
whereas assumption (e) ensures that upon being finalized, these blocks are soon
adopted by the light clients through the light client updates.
Together with (c), these assumptions and Thm.~\ref{thm:security}
imply the security of our \optclient and superlight client
constructions for Ethereum per Def.~\ref{def:state-security}.

\myparagraph{Security under adversarial network conditions.}

Due to network delays or temporary adversarial majorities, there might be extended periods during which the light client
does not receive any updates.
In this case, if the client observes that $\mathsf{UPDATE\_TIMEOUT}$ number of slots have passed since the slot of the
last $\mathsf{finalized\_header}$ in its view, it can do a \emph{force update}.
Prior to the force update, the client replaces the $\mathsf{finalized\_header}$ within the best valid light client update
in its view with the $\mathsf{attested\_header}$ of the same update.
Note that the $\mathsf{finalized\_header}$ of the best valid update must have had a smaller slot than
the $\mathsf{finalized\_header}$ in the client's view, as it could not prompt the client to update its view
during the last $\mathsf{UPDATE\_TIMEOUT}$ slots.
Hence, treating the $\mathsf{attested\_header}$ within the best valid update, which is by definition from a higher slot,
as a $\mathsf{finalized\_header}$ can enable the client to adopt it as the latest
$\mathsf{finalized\_header}$ block, and facilitate the client's progression into a later sync committee period.

The current Ethereum specification~\cite{minimal_light_client} also recommends using other use-case dependent heuristics
for updates, in lieu of checking signatures, if the light client seems stalled.
However, heuristics such as swapping the attested and finalized headers as described above might cause the
light client to adopt block headers that are not finalized by Casper FFG.
Hence, in this work, we assume that the underlying consensus protocol is not subject to disruptions like network delays,
and focus on the regular update mechanism described in Sec.~\ref{sec.pos-eth}.

\section{Discussion and Future Work}

\myparagraph{Interactivity.}
The interactivity in our protocol is undesirable. Contrary to linear bootstrapping protocols in which the
bottleneck in practice comes down to the bandwidth needed to download all the data, in our protocol,
because the communication complexity we have achieved is so low,
the practical bottleneck comes down to the interactivity.
This interactivity is inherent to our construction and is difficult to remove.
In contrast to standard cryptography
proof protocols (such as zero-knowledge proofs), our interactivity cannot be removed using the Fiat--Shamir
heuristic~\cite{fiatshamir}. The reason here is that we require two provers to challenge each other under the watchful
eyes of the verifier. Such behavior cannot be emulated by the random sampling that Fiat--Shamir would offer,
in which a \emph{single} prover and \emph{single} verifier interaction is emulated.
However, even though the protocol has inherent logarithmic interactivity, in practice the actual number
of interaction rounds can be reduced with a couple of techniques:

\begin{enumerate}
    \item The tournament
          between multiple provers can involve bisection games played in parallel. If the verifier speaks to $8$
          provers, once every prover commits to his tree root, the verifier can have the provers compete with
          each other in $4$ bisection games running simultaneously over the network, to reduce latency.
          The $4$ remaining provers can play against each other in $2$ bisection games running in parallel,
          and so forth. Although this does not decrease the number of interactions, the actual network delay
          will be significantly reduced as compared to challenging the parties one-by-one.
\end{enumerate}

Despite the above optimizations, the asymptotic interaction remains $\mathcal{O}(\log n)$.
The problem of \emph{Non-Interactive} Proof of Proof-of-Stake (NIPoPoS), PoPoS that can run in
just a single interaction, is left for future work.

\myparagraph{Trusted setup.}
Our protocol was developed in the standard model and does not require a trusted setup.
The underlying proof-of-stake protocol \emph{does} have some trusted setup assumption:
The genesis epoch contains the public keys of a committee who, at the time of the epoch,
were assumed to have an honest majority. The fact that our protocol does not leverage that
trust to build zero knowledge or other constructions has certain advantages. We only use
hashes and signatures. This makes the protocol easy to understand and straightforward to implement,
with a limited attack surface, a great advantage in security-critical protocols. But also,
in case the trusted setup assumption is violated, our protocol can readily provide evidence
of what exactly happened, because it does not rely on that assumption: At the exact point
where two provers disagree at the bisection game, the evidence presented illustrates two
different epochs at the same index accompanied by their different randomnesses, a situation
that should not occur. In contrast,
a zero-knowledge proof in case of trusted setup failure can lead to decisions in which
foul play may be undetectable, and certainly the exact point and conditions of first
failure are difficult to pinpoint. Additionally, the simplicity of our construction makes
it possible to implement very efficient provers and verifiers in practice. Despite the
slightly worse asymptotics, we expect our construction to perform better in concrete
terms than zero knowledge proof systems that feature large constants in their complexities.

\myparagraph{Proof-of-Work extensions.}
Our tree-based protocol cannot be readily extended to proof-of-work. To see why, consider
a ``chain tree'' similar to our handover tree, but storing the proof-of-work chain in its leaves.
One could na\"ively imagine that two provers can run a similar bisection protocol to
find the first difference in this tree. But note that, contrary to proof-of-stake epochs,
there can be multiple trees that are \emph{admissible} here, and they are not prefixes of one
another. For example, if a mining adversary chooses to mine a secret chain starting
somewhere in the middle of the honest chain, she will end up with a ``chain tree'' whose first
difference with the honest tree will be the first secret block. At that point, the honest
verifier has nothing to see; both trees look equally valid when the leaves are revealed. Therefore,
this scheme cannot be readily applied to \emph{chains}, but must remain constructed on top of
\emph{epochs}, as we chose to do. The closest related proof-of-work construction is FlyClient~\cite{flyclient}
in which leaves are opened at random until foul play is detected, but no bisection games
are played.

\section{Proofs}\label{sec.proofs}

\begin{proof}[Proof of Thm.~\ref{thm:succinctness}]
    Let $N\in\Theta(r)$ be the number of epochs at round $r$.
    When the handover trees have $N$ leaves,
    there can be at most $\log{N}\in\Theta(\log{r})$ steps of interactivity during the bisection game.
    In case an adversarial prover attempts to continue beyond $\log N$ steps
    of interactivity, the verifier aborts the interaction early,
    as the verifier expects to receive sync committees after $\log{N}$ queries,
    and the number $N$ is known by the verifier.

    At each step of the bisection game until the sync committees are revealed,
    the verifier receives two children (two constant size hash values) of the queried node from both provers.
    At the final step, the verifier receives the sync committees $S^{j}$ and $S^{*,j}$ from the provers at
    the first point of disagreement $j$, and the sync committees $S^{j-1}$ and $S^{*,j-1}$ at the preceding leaf
    along with their Merkle proofs.
    As each committee consists of a constant number $m$ of public keys with constant size,
    and each Merkle proof contains $\log{N}$ constant size hash values,
    the total communication complexity of the bisection game becomes $\Theta(\log{N})=\Theta(\log(r))$.
\end{proof}

\begin{proof}[Proof of Thm.~\ref{thm:completeness}]
    To show that the honest prover wins the bisection game,
    we will step through the conditions checked by the verifier during the bisection game.

    At the start of the game,
    the honest prover and verifier both agree on the number of past epochs $N$.
    By synchrony, the honest prover does not time out and
    replies to all of the \emph{open} queries sent by the verifier.
    As the honest prover's handover tree is well-formed,
    at each \emph{open} query asking the honest prover to reveal the children of a node $h_c$ on its tree,
    the left and the right children $h_l$ and $h_r$ returned by the honest prover satisfy
    the relation $h_c=H(h_l \concat h_r)$.
    Thus, the replies are always syntactically valid and accepted by the verifier.
    Subsequently, upon reaching a leaf, the honest prover supplies
    a sync committee, as expected by the verifier.

    Suppose the first point of disagreement between the leaves of the honest and the
    adversarial prover is identified at some index $j$.
    If $j=0$, the honest prover returns $S^0$, which is validated by the verifier
    as the correct sync committee supplied by the genesis state $\genesisstate$.

    If $j>0$, the honest prover reveals the sync committee $S^j$ at leaf $j$,
    the committee $S^{j-1}$ at leaf $j-1$, and the Merkle inclusion proof for $S^{j-1}$,
    which is validated by the verifier with respect to the root.
    By the well-formedness of the honest prover's handover tree, the prover holds a handover proof
    $\Sigma^j$ that contains over $m/2$ signatures on $(j,S^j)$ by unique committee members within $S^{j-1}$.
    The honest prover sends this valid handover proof to the verifier.
    Consequently, the honest prover passes all of the verifier's checks, and wins the bisection game.
\end{proof}

Let $\textsc{Verify}$ be the verification function for Merkle proofs.
It takes a proof $\pi$, a Merkle root $\mroot$, the size of the tree $\ell$,
an index for the leaf $0 \leq i < \ell$ and the leaf $v$ itself.
It outputs $1$ if $\pi$ is valid and $0$ otherwise.
We assume that the well-formed Merkle trees built with a
collision-resistant hash function satisfy the following collision-resistance property:
\begin{proposition}[Merkle Security~\cite{lazylight}]
    \label{prop:adv-merkle}
    Let $H^s$ be a collision resistant hash function used in the binary Merkle trees.
    For all PPT $\mathcal{A}$:
    $
        \Pr[(v,D,\pi,i) \gets \mathcal{A}(1^\lambda): \mroot = \textsc{MakeMT}(D).\mathrm{root} \land
            D[i] \neq v \land \textsc{Verify}(\pi, \mroot, |D|, i, v) = 1] \leq \text{negl}(\lambda)
    $.
\end{proposition}

The following lemma shows that the sync committees
at the first point of disagreement identified by the verifier are different,
and the committees at the previous leaf are the same with overwhelming probability.

\begin{lemma}[Bisection Pinpointing]
    \label{lem:bisection.pinpointing}
    Let $H^s$ be a collision resistant hash function.
    Consider the following game among an honest prover $P$,
    a verifier $V$ and an adversarial prover $P^*$:
    The prover $P$ receives an array $D$ of size $N$ from $P^*$,
    and calculates the corresponding Merkle tree $\mathcal{T}$ with root $\mroot$.
    Then, $V$ mediates a bisection game between $P^*$
    claiming root $\mroot^*$ and $P$ with $\mroot$.
    Finally, $V$ outputs $(1, D^*[j-1], D^*[j])$ if $P^*$ wins the bisection game;
    otherwise, it outputs $(0, \bot, \bot)$.
    Here, $D^*[j-1]$ and $D^*[j]$ are the two entries revealed by $P^*$
    for the consecutive indices $j-1$ and $j$ during the bisection game.
    ($D^*[-1]$ is defined as $\bot$ if $j=0$.)
    Then, for all PPT adversarial provers $\mathcal{A}$,
    $\Pr[D \gets \mathcal{A}(1^\lambda);
            (1, D^*[j-1], D^*[j]) \gets (V(|D|) \leftrightarrow (P(D), \mathcal{A}))
            \land (D^*[j-1] \neq D[j-1] \lor D^*[j] = D[j])]
        \leq \negl(\lambda)$.
\end{lemma}

The above lemma resembles~\cite[Lemma 4]{lazylight} and its proof is given below:

\begin{proof}[Proof of Lem.~\ref{lem:bisection.pinpointing}]
    Consider an adversary $\mathcal{A}(1^{\lambda})$ such that
    $(1, D^*[j-1], D^*[j]) \gets (V(|D|) \leftrightarrow (P(D), \mathcal{A}))
        \land (D^*[j-1] \neq D[j-1] \lor D^*[j] = D[j])$.
    We next construct an adversary $\mathcal{A}_m$ that uses $\mathcal{A}$ as a subroutine
    to break Merkle security.

    The verifier starts the bisection game
    by asking the provers to reveal the children of the roots $\mroot$ and $\mroot^*$ of
    the respective handover trees, where $\mroot \neq \mroot^*$.
    Subsequently, at every step of the bisection game, the verifier asks each prover to
    reveal the two children of a previously revealed node, where the queried nodes have
    the same position, yet different values in the respective trees.
    Hence, for the index $j$ identified by the verifier
    as the first point of disagreement, it holds that $D^*[j] \neq D[j]$.
    Since $D^*[-1] = D[-1] = \bot$, for $j=0$,
    $\Pr[D \gets \mathcal{A}(1^\lambda);
            (1, D^*[j-1], D^*[j]) \gets (V(|D|) \leftrightarrow (P(D), \mathcal{A}))
            \land (D^*[j-1] \neq D[j-1] \lor D^*[j] = D[j])]
        = 0$.

    If $j>0$, there exists a step in the bisection game,
    where the verifier asks the provers to open the right child of the previously queried node.
    Concretely, there exists a node $\tilde{h}_c$ on $\mathcal{T}$, queried
    by the verifier, and a node $\tilde{h}^*_c$, alleged by $P^*$ to be
    at the same position as $\tilde{h}_c$, such that for the two children $\tilde{h}_l$ and $\tilde{h}_r$
    of $\tilde{h}_c$ and the two children $\tilde{h}^*_l$ and $\tilde{h}^*_r$
    of $\tilde{h}^*_c$ revealed to the verifier, the following holds: $\tilde{h}^*_l = \tilde{h}_l$ and
    $\tilde{h}^*_r \neq \tilde{h}_r$.
    Let's consider the last such nodes $\tilde{h}^e_c$ and $\tilde{h}^{e,*}_c$ after which,
    the verifier asks the provers to open only the left children of the subsequent nodes.
    Let $\tilde{h}^e_l$ denote the left child of $\tilde{h}^e_c$, which by definition equals
    the left child of $\tilde{h}^{e,*}_c$ alleged by the adversary.
    Let $D'$ denote the sequence of leaves that lie within the subtree $\mathcal{T}'$ rooted at $\tilde{h}^e_l$.
    Note that the honest verifier knows the number of leaves, \ie, $|D'|$, within the subtree $\mathcal{T}'$.

    Consider the Merkle proofs $\pi$ and $\pi^*$ revealed for $D[j-1]$ and $D^*[j-1]$
    with respect to $\mroot$ and $\mroot^*$ respectively.
    Let $b_a, b_{a-1}, \ldots,b_2, b_1$ denote the binary representation of $j-1$ from the most important bit to the least
    (The index of the first leaf is zero).
    Given $a := \log{(|D|)}$, the verifier can parse the Merkle proofs as
    $\pi = (h_1,h_2, \ldots, h_{a})$ and $\pi^* = (h^*_1,h^*_2, \ldots, h^*_{a})$.
    Since $(1, D^*[j-1], D^*[j]) \gets (V(|D|) \leftrightarrow (P(D), \mathcal{A}))$, $\pi^*$ verifies with respect to $\mroot^*$:
    \begin{itemize}
        \item $h^{*,f}_1 := H(D^*[j-1])$.
        \item $h^{*,f}_{a+1} := \mroot^*$.
        \item For $i=1, \ldots, a$; $h^{*,f}_{i+1} := H(h^{*,f}_i,h^*_i)$ if $b_i = 0$,
              and $h^{*,f}_{i+1} := H(h^*_i,h^{*,f}_i)$ if $b_i = 1$.
    \end{itemize}

    Now, consider the prefix of the Merkle proof $\pi^*$ consisting of the first $\log(|D'|)$ entries:
    $\pi^*_p = (h_1, \ldots, h_{\log(|D'|})$.
    By definition of $j$, the indices $b_{\log(|D'|)}, \ldots, b_{1}$ are all $1$, and;
    \begin{itemize}
        \item $h^{*,f}_1 = H(D^*[j-1])$.
        \item $h^{*,f}_{\log(|D'|)+1} = \tilde{h}^e_l$.
        \item For $i=1, \ldots, \log(|D'|)$; $h^{*,f}_{i+1} = H(h^*_i,h^{*,f}_i)$.
    \end{itemize}
    Hence, it holds that $\textsc{Verify}(\pi^*_p, \tilde{h}^e_l, |D'|, |D'|-1, D^*[j-1]) = 1$.
    Moreover, $\tilde{h}^e_l = \textsc{MakeMT}(D').\mathrm{root}$ and
    $D'[|D'|-1] = D[j-1] \neq D^*[j-1]$.

    Finally, $\mathcal{A}_m$ uses $\mathcal{A}$ as a subroutine to generate $D$, $\pi^*$ and $D^*[j-1]$,
    and outputs $(D^*[j-1], D', \pi^*_p, |D'|-1)$, which implies that
    $\tilde{h}^e_l:= \textsc{MakeMT}(D').\mathrm{root}$, $D'[|D'|-1] = D[j-1] \neq D^*[j-1]$ and
    $\textsc{Verify}(\pi^*_p, \tilde{h}^e_l, |D'|, |D'|-1, D^*[j-1]) = 1$.
    Consequently, by Prop.~\ref{prop:adv-merkle}, for all PPT adversarial provers $\mathcal{A}$,
    $\Pr[D \gets \mathcal{A}(1^\lambda);
            (1, D^*[j-1], D^*[j]) \gets (V(|D|) \leftrightarrow (P(D), \mathcal{A}))
            \land (D^*[j-1] \neq D[j-1])] \leq \negl(\lambda)$.
    As $D^*[j] \neq D[j]$, this implies that for all PPT adversarial provers $\mathcal{A}$,
    $\Pr[D \gets \mathcal{A}(1^\lambda);
            (1, D^*[j-1], D^*[j]) \gets (V(|D|) \leftrightarrow (P(D), \mathcal{A}))
            \land (D^*[j-1] \neq D[j-1] \lor D^*[j] = D[j])]
        \leq \negl(\lambda)$.
\end{proof}

\begin{definition}[Definition 13.1 of Boneh \& Shoup~\cite{crypto-book-boneh}]
    \label{def:signature-scheme}
    A signature scheme $\mathbf{S} = (G, S, V)$ is a triple of efficient algorithms, $G$, $S$ and
    $V$, where $G$ is called a key generation algorithm, $S$ is called a signing algorithm, and $V$ is
    called a verification algorithm. Algorithm $S$ is used to generate signatures and algorithm $V$ is
    used to verify signatures.
    \begin{itemize}
        \item $G$ is a probabilistic algorithm that takes no input.
              It outputs a pair $(pk, sk)$, where $sk$ is called a secret signing key and $pk$ is called a public verification key.
        \item $S$ is a probabilistic algorithm that is invoked as $\sigma \xleftarrow{R}{} S(sk, m)$, where $sk$ is a secret key
              (as output by $G$) and $m$ is a message. The algorithm outputs a signature $\sigma$.
        \item $V$ is a deterministic algorithm invoked as $V(pk, m, \sigma)$. It outputs either accept or reject.
        \item We require that a signature generated by $S$ is always accepted by $V$ (valid for short). That is, for all $(pk, sk)$
              output by $G$ and all messages $m$, we have $\Pr[V(pk, m, S(sk, m)) = \mathrm{accept}] = 1$.
    \end{itemize}
    We say that messages lie in a finite message space $\mathcal{M}$,
    signatures lie in some finite signature space $\Sigma$, and $\mathbf{S} = (G, S, V)$ is defined over $(\mathcal{M}, \Sigma)$.
\end{definition}

\begin{definition}[Attack Game 13.1 of Boneh \& Shoup~\cite{crypto-book-boneh}]
    \label{def:attack-game}
    For a given signature scheme $\mathbf{S} = (G, S, V)$, defined
    over $(\mathcal{M}, \Sigma)$, and a given adversary $\mathcal{A}$, the attack game runs as follows:
    \begin{itemize}
        \item The challenger runs $(pk, sk) \xleftarrow{R}{} G()$ and sends $pk$ to $\mathcal{A}$.
        \item $\mathcal{A}$ queries the challenger several times.
              For $i = 1, 2, \ldots$, the $i^{\text{th}}$ signing query is a message $m_i \in \mathcal{M}$.
              Given $m_i$, the challenger computes $\sigma_i \xleftarrow{R}{} S(sk, m_i)$, and then gives $\sigma_i$ to $\mathcal{A}$.
        \item Eventually $\mathcal{A}$ outputs a candidate forgery pair $(m, \sigma) \in \mathcal{M} \times \Sigma$.
    \end{itemize}
    We say that the adversary wins the game if the following two conditions hold:
    \begin{itemize}
        \item $V(pk, m, \sigma) = \mathrm{accept}$, and
        \item $m$ is new, namely $m \notin \{m_1, m_2, \ldots \}$.
    \end{itemize}
    We define $\mathcal{A}$’s advantage with respect to $\mathbf{S}$, denoted by $\mathrm{SIGadv}[\mathcal{A}, \mathbf{S}]$,
    as the probability that $\mathcal{A}$ wins the game.
\end{definition}

\begin{definition}[Definition 13.2 of Boneh \& Shoup~\cite{crypto-book-boneh}]
    \label{def:existential-unforgeability}
    We say that a signature scheme $\mathbf{S}$ satisfies existential unforgeability under a chosen message attack
    (existential unforgeability for short) if for all efficient adversaries $\mathcal{A}$,
    the quantity $\mathrm{SIGadv}[\mathcal{A}, \mathbf{S}]$ is negligible.
\end{definition}

\begin{proof}[Proof of Thm.~\ref{thm:soundness}]
    Consider the following game among an honest prover $P$,
    a verifier $V$ and an adversarial prover $\mathcal{A}$:
    The prover $P$ receives an array $D = (S^0, \ldots, S^{N-1})$ of sync committees from the underlying PoS protocol,
    and calculates the corresponding Merkle tree $\mathcal{T}$ with root $\mroot$.
    Similarly, the prover $P$ receives a succession of handover proofs
    $\mathbb{S} = (\Sigma^1, \Sigma^2, \ldots, \Sigma^{N-1})$,
    where for all $j=1, \ldots, N-1$, $\Sigma^j$ consists of over $m/2$ valid signatures on
    $(j+1, S^{j+1})$ by unique honest sync committee members assigned to epoch $j$.
    Then, $V$ mediates a bisection game between $\mathcal{A}$
    claiming root $\mroot^*$ and $P$ claiming root $\mroot$.
    Finally, $\mathcal{A}$ wins the bisection game.
    In the subsequent proof, we will construct an adversary $\mathcal{A}_s$
    that uses $\mathcal{A}$ as a subroutine to break the existential unforgeability of the signature scheme
    under a chosen message attack.

    Let $j$ denote the first point of disagreement between the leaves of the honest and the
    adversarial provers $P$ and $\mathcal{A}$.
    If $j=0$, let $S^0$ and $S^{*,0}$ denote the committees returned by the honest and
    adversarial provers respectively for the first leaf.
    By Lem.~\ref{lem:bisection.pinpointing}, $S^0 \neq S^{*,0}$.
    As the honest prover's tree is well-formed,
    $S^0$ is the sync committee within the genesis state $\genesisstate$.
    Thus, in this case, $\mathcal{A}$ loses the bisection game, which implies $j>0$.

    During epoch $j-1$, the honest committee members assigned to epoch $j-1$ constitute
    over $m/2$ of the members within $S^{j-1}$, and create only a \emph{single} handover signature on $(j,\Sigma^{j})$.
    After epoch $j-1$ ends, no PPT adversary can access the secret signing keys
    of the honest members of the committee $S^{j-1}$ due to the use of key-evolving signatures.

    Let $S^{*,j-1}$ and $S^{*,j}$ denote the sync committees revealed by $\mathcal{A}$
    for the consecutive indices $j-1$ and $j$ during the bisection game.
    Suppose $S^j \neq S^{*,j}$ and $S^{j-1} = S^{*,j-1}$.
    Since $\mathcal{A}$ wins the bisection game, it provides a handover proof $\Sigma^{*,j}$
    that contains over $m/2$ signatures on $(j,S^{*,j})$ by unique committee members within $S^{*,j-1} = S^{j-1}$.
    Thus, there exists at least one committee member $S^{j-1}_{i^*}$ with the smallest index such that
    \begin{itemize}
        \item There is a signature $\sigma^{*,j-1}_{i^*}$ within the handover proof $\Sigma^{*,j}$
              such that given the public verification key $pk$ of $S^{j-1}_{i^*}$, $V(pk, (j,S^{*,j}), \sigma^{*,j-1}_{i^*}) = \mathrm{accept}$.
        \item During epoch $j-1$, $S^{j-1}_{i^*}$ was an honest committee member assigned to epoch $j-1$.
        \item $S^{j-1}_{i^*}$ has created only a single handover signature $\sigma^{j-1}_{i^*}$ on $(j, S^j)$ during epoch $j-1$.
        \item After epoch $j-1$ ends, no PPT adversary can access the secret signing key of $S^{j-1}_{i^*}$.
    \end{itemize}
    Consequently, $\mathcal{A}$ provides a signature $\sigma^{*,j-1}_{i^*}$ on $(j,S^{*,j})$ that verifies with respect to the
    public verification key of $S^{j-1}_{i^*}$.

    Given the array of sync committees $D = (S^0, \ldots, S^{N-1})$ from the underlying PoS protocol,
    we next construct an existential forgery adversary $\mathcal{A}_s$ that has access
    to the adversarial prover $\mathcal{A}$ as a subroutine.
    During $\mathcal{A}_s$'s interaction with $\mathcal{A}$,
    it receives signing queries from the adversarial and honest sync committee members within $S_0 \ldots S_{N-1}$,
    and passes these queries to the challenger, which replies with the queried signatures.
    It then passes the signatures back to $\mathcal{A}$, and
    the succession of handover proofs $\mathbb{S} = (\Sigma^1, \Sigma^2, \ldots, \Sigma^{N-1})$ to
    the honest prover $P$ as specified at the beginning of the proof.
    Finally, $\mathcal{A}_s$ obtains the handover proofs $\Sigma^j$ and $\Sigma^{*,j}$ from $\mathcal{A}$,
    and identifies $S^{j-1}_{i^*}$.
    It subsequently outputs $\sigma^{*,j-1}_{i^*}$ on the message $(j,S^{*,j})$, for which the following conditions hold:
    \begin{itemize}
        \item Given the public verification key $pk$ of $S^{j-1}_{i^*}$, it holds that
              $V(pk, (j,S^{*,j}), \sigma^{*,j-1}_{i^*}) = \mathrm{accept}$, and
        \item $(j,S^{*,j}) \neq (j,S^j)$, where unlike $(j,S^j)$,
              the message $(j,S^{*,j})$ was not sent as a query to the challenger.
    \end{itemize}
    Thus, $\mathcal{A}_s$ wins the attack game in Def.~\ref{def:attack-game}.

    Finally, if there is a PPT adversary $\mathcal{A}$ such that $\mathcal{A}$ wins the bisection game
    against the honest prover and the sync committees received by the verifier satisfies
    $S^j \neq S^{*,j}$ and $S^{j-1} = S^{*,j-1}$, $\mathcal{A}_s$ described above wins the attack game.
    By Lem.~\ref{lem:bisection.pinpointing}, for all PPT adversaries $\mathcal{A}$,
    $S^j \neq S^{*,j}$ and $S^{j-1} = S^{*,j-1}$ with overwhelming probability.
    Moreover, as the signature scheme satisfies existential unforgeability,
    for all PPT adversaries $\mathcal{A}$, the adversary $\mathcal{A}$ loses the attack game
    with overwhelming probability.
    Consequently, for all PPT adversarial provers $\mathcal{A}$, the prover $\mathcal{A}$ loses the bisection game
    against the honest prover with overwhelming probability in $\lambda$.
\end{proof}

\begin{proof}[Proof of Thm.~\ref{thm:tournament-runtime}]
    Consider a tournament started at round $r$ with $|\mathcal{P}|$ provers,
    one of which is honest.
    At each step of the tournament, the verifier facilitates a bisection game
    between two provers with different state commitments.
    (Honest provers hold the same state commitment.)
    At the end of the game, at least one prover is designated as a loser
    and eliminated from the set of provers.
    The tournament continues until all remaining provers hold the same state commitment.
    Hence, it lasts at most $|\mathcal{P}|-1$ steps.
    By Thm.~\ref{thm:succinctness}, each bisection game at round $r$
    ends in $O(\log(r))$ steps of interactivity, and
    has a total communication complexity of $O(\log(r))$.
    Consequently, the tournament consists of $O(|\mathcal{P}|\log(r))$ steps of interactivity,
    and has a total communication complexity of $O(|\mathcal{P}|\log(r))$.
\end{proof}

\begin{proof}[Proof of Thm.~\ref{thm:security}]
    Consider a tournament step that involves an honest prover $P$ and an adversarial prover $P^*$
    that have provided different state commitments $\stc$ and $\stc^*$ respectively, for the state of
    the blockchain at the beginning of the epoch containing round $r$.
    Let $N$ denote the number of past epochs at round $r$ (starting at epoch $0$), and
    $S^{N-1}$ denote the committee assigned to epoch $N-1$.
    Define $n$ as the number of Merkle trees within the MMRs of the honest provers at epoch $N$,
    and let $D_i$ denote the sequence of leaves within the $i^{\text{th}}$ tree of the honest prover.
    Let $\mroot^*_i$ and $\mroot_i$, $i \in [n]$, denote the sequence of peaks revealed by $P^*$
    and $P$ to the verifier before the bisection game.
    By definition, $P$ returns $S^{N-1}$ as the latest sync committee in its view, and
    let $S^{*,N-1}$ denote the latest sync committee alleged by $P^*$.
    The prover $P$ sends over $m/2$ signatures on $\stc$ by unique committee members within $S^{N-1}$, whereas
    $P^*$ sends over $m/2$ signatures on $\stc^*$ by unique committee members within $S^{*,N-1}$.
    Similarly, the prover $P$ sends a Merkle proof $\pi$ such that
    $\textsc{Verify}(\pi, \mroot_n, |D_n|, |D_n|-1, S^{N-1}) = 1$, whereas $P^*$ sends a Merkle proof $\pi^*$ such that
    $\textsc{Verify}(\pi^*, \mroot^*_n, |D_n|, |D_n|-1, S^{*,N-1}) = 1$.
    We first show that $S^{*,N-1} \neq S^{N-1}$ with overwhelming probability.
    We will then prove that if $S^{*,N-1} \neq S^{N-1}$, then $\mroot^*_n \neq \mroot_n$, with overwhelming probability.

    To show that $S^{*,N-1} \neq S^{N-1}$, we construct an existential forgery adversary $\mathcal{A}_s$
    that uses the adversarial prover $P^*$ as a subroutine to break the existential unforgeability of the signature scheme
    under a chosen message attack.
    Suppose $S^{*,N-1} = S^{N-1}$.
    At the beginning of epoch $N-1$, the honest committee members assigned to epoch $N-1$ constitute
    over $m/2$ of the members within $S^{N-1}$, and create only a \emph{single} signature on a state commitment,
    namely $\stc$.
    Since $P^*$ sends over $m/2$ signatures on $\stc^*$ by unique committee members within $S^{N-1}$,
    there is at least one committee member $S^{N-1}_{i^*}$ within $S^{N-1}$ with the smallest index such that
    \begin{itemize}
        \item There is a signature $\sigma^*$ such that given the public verification key $pk$ of
              $S^{N-1}_{i^*}$, it holds that $V(pk, \stc^*, \sigma^*) = \mathrm{accept}$.
        \item During epoch $N-1$, $S^{N-1}_{i^*}$ is an honest committee member assigned to epoch $N-1$.
        \item $S^{N-1}_{i^*}$ has created only a single signature $\sigma$ on a state commitment during epoch $N-1$,
              and that is on $\stc$.
    \end{itemize}
    Consequently, $P^*$ provides a signature $\sigma^*$ on $\stc^*$ that verifies with respect to the
    public verification key of $S^{N-1}_{i^*}$.

    We next construct the adversary $\mathcal{A}_s$ that has access
    to the adversarial prover $P^*$ as a subroutine.
    During $\mathcal{A}_s$'s interaction with $P^*$,
    it receives signing queries on state commitments
    from the adversarial and honest sync committee members within $S_{N-1}$,
    and passes these queries to the challenger, which replies with the queried signatures.
    It then passes the signatures back to $P^*$ and $P$.
    Finally, $\mathcal{A}_s$ obtains $m/2$ signatures on the commitments $\stc$ and $\stc^*$ from $\mathcal{A}$,
    and identifies $S^{N-1}_{i^*}$.
    It subsequently outputs $\sigma^*$ on the message $\stc^*$, for which the following conditions hold:
    \begin{itemize}
        \item Given the public verification key $pk$ of $S^{N-1}_{i^*}$, it holds that
              $V(pk, \stc^*, \sigma^*) = \mathrm{accept}$, and
        \item $\stc^* \neq \stc$, where unlike $\stc$,
              the message $\stc^*$ was not sent as part of a signing query to the challenger.
    \end{itemize}
    Thus, $\mathcal{A}_s$ wins the attack game in Def.~\ref{def:attack-game}.

    Finally, if there is a PPT adversary $P^*$ such that it gives $m/2$ signatures on $\stc^*$ by unique committee members
    within $S^{N-1}$, $\mathcal{A}_s$ described above wins the attack game.
    However, as the signature scheme satisfies existential unforgeability,
    for all PPT adversaries $\mathcal{A}$, the adversary $\mathcal{A}$ loses the attack game
    with overwhelming probability.
    Consequently, for all PPT adversarial provers $P^*$, it holds that $S^{*,N-1} \neq S^{N-1}$ with overwhelming probability.

    Next, we show that if $S^{*,N-1} \neq S^{N-1}$, then $\mroot^*_n \neq \mroot_n$ with overwhelming probability.
    Towards this goal, we construct an adversary $\mathcal{A}_m$ that uses $P^*$ as a subroutine to break Merkle security.
    Suppose $S^{*,N-1} \neq S^{N-1}$ and $\mroot^*_n = \mroot_n$.
    In this case, $\mathcal{A}_m$ receives from $P^*$, the set $S^{*,N-1}$, the sequence of leaves $D_n$ within the last
    tree of the honest prover's MMR, the proof $\pi^*$ and the index $|D_n|-1$.
    It then outputs $(S^{*,N-1}, D_n, \pi^*, |D_n|-1)$, for which it holds that
    $D_n[|D_n|-1] = S^{N-1} \neq S^{*,N-1}$ and
    $\textsc{Verify}(\pi^*, \mroot^*_n, |D_n|, |D_n|-1, S^{*,N-1})
        = \textsc{Verify}(\pi^*, \mroot_n, |D_n|, |D_n|-1, S^{*,N-1}) = 1$, where $\mroot_n$ is the root of the last
    Merkle tree (that has the leaves $D_n$) within the honest prover's MMR.
    However, by Prop.~\ref{prop:adv-merkle}, we know that for all PPT adversaries $\mathcal{A}$:
    $
        \Pr[(v,D,\pi,i) \gets \mathcal{A}(1^\lambda): \mroot = \textsc{MakeMT}(D).\mathrm{root} \land
            D[i] \neq v \land \textsc{Verify}(\pi, \mroot, |D|, i, v) = 1] \leq \text{negl}(\lambda)
    $.
    Hence, for all PPT adversarial provers $P^*$ with state commitment $\stc^* \neq \stc$, $S^{*,N-1} \neq S^{N-1}$, and
    the sequence of peaks $\mroot^*_i$ revealed to the verifier by the
    adversarial prover is different from the sequence $\mroot_i$, $i \in [n]$, revealed by the honest prover with overwhelming
    probability.
    Thus, there exists an index $d \in [n]$ such that $\mroot^*_d \neq \mroot_d$ and $\mroot^*_{i} = \mroot_{i}$
    for all $i \in [n]$, $i<d$, with overwhelming probability.
    In this case, the verifier mediates a bisection game between $P$ and $P^*$
    on the two alleged trees with the roots $\mroot^*_d$ and $\mroot_d$.
    By Thm.~\ref{thm:completeness}, $P$ wins the game, and by Thm.~\ref{thm:soundness}, $P^*$ loses
    the game with overwhelming probability.
    As a result, $P^*$ is eliminated at this tournament step with overwhelming probability.

    At each step of the tournament, at least one prover is eliminated, and
    the tournament continues until all remaining provers hold the same state commitment, with overwhelming probability.
    By assumption, there is at least one honest prover $P$.
    This prover emerges victorious from every tournament step against other provers with a different state commitment,
    except with negligible probability.
    Consequently, with overwhelming probability, $P$ remains in the tournament until all remaining provers hold the same
    state commitment $\stc$ as $P$.

    Let $\ledger$ be the ledger held by $P$ at round $r_0$ corresponding to the beginning of the epoch of round $r$.
    By definition, $r_0 \leq r$ and $r-r_0 \leq C$ for some constant epoch length $C$.
    By the safety of the PoS protocol,
    for any honest parties $P_1, P_2$ and rounds $r_1 \geq r_2$: $\ledger^{P_2}_{r_2} \preccurlyeq \ledger^{P_1}_{r_1}$.
    Thus, for any honest party $P'$ and rounds $r' \geq r \geq r_0$, it holds that $\ledger \preccurlyeq \ledger^{P'}_{r'}$,
    Similarly, for any honest party $P'$, it holds that $\ledger^{P'}_{r_0-1} \preccurlyeq \ledger$.
    Consequently, there exists a latency parameter $\nu=K$,
    and a ledger $\ledger$ such that $\stc = \transition^*(\genesisstate, \ledger)$,
    and $\ledger$ satisfies the following properties:
    \begin{itemize}
        \item \textbf{Safety:} For all rounds $r' \geq r + \nu$: $\ledger \preccurlyeq \ledger^{\cup}_{r'}$.
        \item \textbf{Liveness:} For all rounds $r' \leq r - \nu$: $\ledger^{\cap}_{r'} \preccurlyeq \ledger$.
    \end{itemize}
    Thus, $\stc$ satisfies state security.
    As the verifier accepts $\stc$ as the correct commitment at the end of the tournament
    with overwhelming probability,
    the commitment obtained by the verifier at the end of the tournament satisfies
    state security with overwhelming probability in $\lambda$.
\end{proof}

\myparagraph{Safety under partial synchrony.}
We observe the assumption requiring at least one prover to be honest is essential for both
the safety and liveness of our superlight client. Similarly 
proof of stake protocols without finality (e.g., Ouroboros, Ouroboros Praos) also lose safety when 
eclipsed. In contrast, proof of stake protocols that
have finality (e.g., Ethereum's Casper FFG, Algorand, Tendermint, Streamlet, and HotStuff) 
maintain safety and only lose liveness when eclipsed.  

\section{Other Proof-of-Stake Systems}
\label{sec.other}

We have presented our construction in a generic PoS model, and instantiated it
concretely for Ethereum PoS.
Our construction
is quite general and can be adopted to virtually any PoS
system. Many
PoS
systems are split
into (potentially smaller) epochs in which some sampling from the underlying stake distribution
is performed according to some random number. The random number generation can be performed in multiple
ways. For example, all of Ouroboros~\cite{ouroboros}, Ouroboros Praos~\cite{praos}, and Ouroboros Genesis~\cite{genesis} use a verifiable secret
sharing mechanism, while
Algorand~\cite{algorand} uses a multiparty computation. The stake distribution from which the sampling is performed
could also have various nuances such as delegation, might require locking up one's funds, may exclude
people with very small stake, or may give different weights to different stake ownership. In all
of these cases, a frozen stake distribution from which the final sampling is performed is determined.

Our scheme can be generalized to any
PoS
scheme in which the leader can be verified
from a frozen stake distribution and some randomness, no matter how it is generated, as
long as the block associated with a particular slot can be uniquely determined after it stabilizes
(a property that follows in any blockchain system as long as it observes the common prefix property).
In the scheme we described throughout the paper, the sequence of signatures $\overline{\sigma}^{j+1}$ that are generated
in an epoch $j$ and vouch for the leaders of the next epoch sign the public key set $S^{j+1}$
of the next epoch. To generalize our scheme to any PoS system with randomness and a stake distribution,
the signatures $S^{j+1}$ need not sign the public key sequence any more; instead, they
can sign:

\begin{enumerate}
    \item the epoch randomness $\eta^{j+1}$ of the next epoch, and
    \item the frozen stake distribution $\textsf{SD}^j$ of the current epoch
          that will be used for sampling during the next epoch.
\end{enumerate}

Of course, in
such a scheme, a succinctness problem arises: The stake distribution $\textsf{SD}^j$ might be large.
However, this problem can be overcome by organizing the stake distribution $\textsf{SD}^j$ into a
Merkle tree. This Merkle tree contains one leaf for every satoshi (the smallest cryptocurrency
denomination). The leaf's value is the public key who owns this satoshi. When sampling from
$\textsf{SD}^j$ according to the randomness $\eta^{j+1}$, the prover can provide a proof
that the correct leader was the one that happened to be elected by opening the particular Merkle
tree path at a particular index. That way, the verifier can deduce the last slot leaders of each epoch. Because
the number of satoshis can be large, this Merkle tree can have a large (potentially an exponential)
number of leaves.
However, its root and proofs can be efficiently computed using sparse Merkle tree techniques~\cite{sparse-mt}
(or Merkle tries~\cite{wood})
because the tree contains a polynomial number of continuous ranges in which many consecutive
leaves share the same value.
Even better, Merkle--Segment trees~\cite{flyclient}) can be used. These trees are similar to
Merkle trees, except that each node (internal or leaf) is also annotated with a numerical value,
here the total stake under the subtree rooted at the particular node. Each internal node has
the property that its annotated value is the sum of the annotated values of its children.

The above technique is quite generic, but each system has its nuances that must be accounted for.

\myparagraph{Ouroboros/Cardano.}
Our construction can be implemented in Cardano/Ouroboros~\cite{ouroboros}
as presented by electing a committee, but
the underlying longest chain rule lends itself to better implementations for committee
election and signature inclusion.
One way to make use of the Cardano protocol is to extend the epoch duration $R$
by $2k$ slots. In this manner, the randomness and leaders of the next epoch are known
during the last $2k$ slots of the previous epoch (in the vanilla Cardano protocol, the
leaders and randomness of the next epoch only become known at the end of the epoch).
The last $2k$ slots of each epoch are then used to determine the sync committee.
The committee is the leaders of these $2k$ slots, and no separate process is required
to elect it. In each of the last $2k$ slots of an epoch, we add the extra requirement
to the block validity rules that the block producer has included a handover signature
to the correct next epoch committee; otherwise the block is rejected as invalid by
full nodes.

These small changes mean that the Ouroboros protocol can be used almost as-is to support
our PoPoS and do not require any additional mechanisms for electing committees or any
off-chain mechanism for exchanging committee succession signatures, as the blocks themselves
are used as carriers of this information. The critical property of the Ouroboros protocol
that allows us to prove security in this setting is the following lemma:

\begin{lemma}[Honest Subsequence]
    \label{lem.subsequence}
    Consider any continuous window of $2k$ slots within an epoch. If \emph{any}
    $k+1$ keys among these $2k$ are chosen, then at least one of them is guaranteed to
    be honest, except with negligible probability in $k$.
\end{lemma}

Using the above lemma, we see that the last $2k$ slots will necessarily contain $k+1$
honest leaders who will produce correct committee signatures, and so our PoPoS assumption
that the committee has honest majority during its epoch is satisfied. The security of
the protocol then follows from Thm.~\ref{thm:security}.

\myparagraph{Ouroboros Praos and Genesis.}
These two protocols have some similarities to Ouroboros, but also significant differences.
As Ouroboros Praos and Genesis are designed to be resilient to fully
adaptive adversaries, the actual slot leader of each slot is not known \emph{a priori}.
However, a party can himself determine whether he is eligible to be the
slot leader by evaluating a VRF on the epoch randomness and the current slot index
using his private key.
If the VRF output is below a certain threshold, determined by the candidate leader's
stake, then the party is elegible to be a leader for this slot.
The party's public key can then be used by others to verify a proof that the VRF computation
was correct, and that he is indeed a rightful leader.

Because we cannot determine the leaders of the $j+1^\text{st}$ epoch at the end of epoch
$j$, we cannot hope to have the leaders of the $j^\text{th}$ epoch sign off the public
keys of the leaders of epoch $j+1$. However, the above technique,
in which signatures sign the randomness
and a Merkle--Segment tree of the stake distribution, together with the VRF proof, suffices.
In this construction, the signatures of epoch $j$ sign off the randomness for epoch $j+1$ and stake
distribution Merkle tree for epoch $j$. At a later time, when it is revealed who is leader,
the honest prover can provide the VRF proof to the verifier, and the verifier can check
that the leader was indeed rightful. To obtain the VRF threshold, the prover can open the
Merkle--Segment tree to the depth required to illustrate the total sum of the stake of the leader.
Once the leader's stake is revealed, the threshold used in the VRF inequality is validated.

These protocols have several advantages, including security in the semi-synchronous
setting as well as resilience to adaptive adversaries~\cite{praos,genesis}.

\myparagraph{Snow White.}
This protocol uses epochs and every epoch contains a randomness and a
stake distribution from which leaders are sampled~\cite{snowwhite}.
Therefore, our protocol can be readily
adapted to it.

\myparagraph{Algorand.}
Contrary to Ouroboros, Algorand offers immediate finality~\cite{algorand}.
Once a block is broadcast,
any transactions contained within are confirmed and can no longer be reverted. In other
words, its common prefix property holds with a parameter of $k = 1$. To achieve this,
Algorand runs a full Byzantine Agreement protocol for the generation of every block
before moving to the next block. One way to look at it is to think of Algorand as a
coin in which the epoch duration is $R = 1$. Our construction can therefore create a
handover tree in which the leaves are exactly the blocks in the Algorand chain. The
Algorand private sortition mechanism can be used to elect a committee large enough
to ensure honest supermajority (a property required for Algorand's security).
This committee, whose members can be placed in increasing order by their public
key to ensure determinism, can then be used in place of our sequence of public keys,
to sign off the results of the next block. Even though our handover
tree now becomes slightly larger, with its number of leaves equal to the
chain length $|\chain|$, our protocol is still $\mathcal{O}(\log|\chain|)$.

\fi

\end{document}